\documentclass[letterpaper,twocolumn,prb,english,showpacs,preprintnumbers,amsmath,amssymb]{revtex4-1}
\usepackage[latin9]{inputenc}
\usepackage{float}
\usepackage{amsmath}
\usepackage{amssymb}
\usepackage{stmaryrd}
\usepackage{graphicx}
\usepackage{color}
\makeatletter

\providecommand{\tabularnewline}{\\}

\usepackage{epsfig}\usepackage{subfigure}\usepackage{dcolumn}\usepackage{mathrsfs}\usepackage{pifont}\usepackage{amsthm}
\usepackage{bm}\usepackage{color}\usepackage{latexsym}\usepackage{amsfonts}

\usepackage{babel}

\makeatother

\usepackage{babel}
\begin{document}
\title{Quantum coherent electron transport in silicon quantum dots}
\author{Xinyu Zhao}
\author{Xuedong Hu}
\email{xhu@buffalo.edu}

\affiliation{Department of Physics, University at Buffalo, SUNY, Buffalo, New York
14260, USA}
\begin{abstract}
With silicon being the go-to material for spin qubits, and motivated by the demand of a scalable quantum computer architecture for fast and reliable quantum information transfer on-chip, we study coherent electron transport in a silicon double quantum dot.  We first examine the valley-orbital dynamics in a silicon double dot, and discuss how to properly measure the tunnel couplings as well as the valley phase difference between two quantum dots.
We then focus on possible phase and spin flip errors during spin transport across a silicon double dot.  In particular, we clarify correction on the effective $g$-factor for the electron spin from the double dot confinement potential, and quantify the resulting phase error.  We then study spin fidelity loss due to spin-valley mixing, which is a unique feature of silicon quantum dots.  We show that a small phase correction between valleys can cause a significant coherence loss.  We also investigate spin flip errors caused by either an external inhomogeneous magnetic field or the intrinsic spin-orbit coupling.  We show that the presence of valleys makes it possible to have much broader (in terms of interdot detuning) level anti-crossings compared to typical anti-crossings in, for example, a GaAs double dot, and such broad anti-crossings lead to amplification of spin flip errors.  Lastly, we design a pulse sequence to suppress various possible spin flip errors by taking advantage of the multiple level anti-crossings in a silicon double dot and employing Landau-Zener transitions.
\end{abstract}
\pacs{73.63-b, 72.25.Rb, 03.67.Hk.}
\maketitle

\section{\label{sec:Introduction}Introduction}

Spin qubits in silicon quantum dots (SiQD) have promising potentials for quantum information processing due to their long coherence time, helped by isotopic purification that suppresses magnetic noise from nuclear spins \citep{tyryshkin2006JoPCM, bluhm2011NatPhys, petersson2012Nature, pla2012Nature, tyryshkin2012NatMat, Saeedi2013Science, Veldhorst2014NatNano, Sigillito2015PRL}.  Spin relaxation is also generally very slow in Si because of the relatively weak spin-orbit coupling (SOC) \citep{Zwanenburg2013RMP}. Looking to the future, the sophisticated technologies from the silicon industry could potentially provide a powerful boost to the scale-up of silicon qubit architectures \citep{pla2012Nature, tyryshkin2006JoPCM, kawakami2014NatNano, zajac2016PRapplied}.  Indeed, the current commercial 14nm/10nm process technology \citep{chau2003DRC, franklin2012NanoLett}
is already at the level of feature sizes required for gated Si quantum dots. These advantages make SiQD an appealing candidate as a building
block for future semiconductor quantum computers \citep{Zwanenburg2013RMP,pla2013Nature,Saeedi2013Science}.

However, SiQD system does have its own challenges, especially in a conduction band that has multiple minima (valleys) \citep{friesen2006APL, friesen2010PRB, rohling2012NJP, veldhorst2015Nature, zajac2015APL, zajac2016PRapplied, culcer2010PRB, culcer2010PRB2, gamble2016APL, schoenfield2017NatComm}.  A small energy splitting between valley eigenstates can introduce unwanted orbital dynamics to a spin qubit, and spin-valley (SV) coupling can lead to mixture between spin and valley degrees of freedom and cause significant spin relaxation under certain conditions (spin hot spots) \citep{yang2013NatComm}.  Furthermore, in a SiQD, valley-orbit coupling is generally a complex parameter. Its magnitude determines the size of the valley splitting, and has been widely studied in a quantum dot \citep{friesen2007PRB, friesen2006APL, wu2012PRB, boross2016PRB}.  Its phase
does not lead to any observable effect in a single dot.  However, in a double dot, the valley-orbit phase difference between the two dots is of great importance in determining interdot tunneling and exchange coupling \citep{wu2012PRB}, but has yet been studied thoroughly.  The availability of low-energy excited valley states usually means a more complex electron spin-orbital dynamics, which could be potentially useful for coherent
manipulation, but may also lead to enhanced relaxation hot spots because of state mixing. As such a quantum information processor based on spin qubits in Si requires a careful examination of the coupled spin-orbit-valley dynamics in the context of high-fidelity coherent manipulations in coupled Si quantum dots \citep{bauerle2018RPP}.

The ability to quickly and reliably distribute information is crucial to a scaled-up quantum computer \citep{Sanada2011PRL, McNeil2011Nature, Huang2013PRB, bluhm2011NatPhys, Zhao2016SR}.  In this context, quantum coherent electron transport over multiple QDs could be one of the fundamental operations for a spin-qubit based quantum information processor \citep{lu2003Nature, li2010PRB, petta2004PRL, veldhorst2015Nature, zajac2015APL, bluhm2011NatPhys, bauerle2018RPP, hermelin2011Nature, flentje2017APL, bertrand2016NatNano, bertrand2016Nano, li2017PRA, mills2018arXiv}.
The aim is to transport an electron over a finite distance without disturbing its spin state in which quantum information is encoded \citep{greentree2004PRB, lu2003Nature, wang2013PRL, hermelin2013JAP, bertrand2016NatNano, hermelin2017PSSb, flentje2017NC, Zhao2018SR}.  Other important tasks such as quantum error correction and quantum measurement may also involve electron tunneling between quantum dots
\citep{OGorman2016NPJQI, hill2015SciAdv, greentree2004PRB, VanHouten1996PhysToday, bertrand2015PRL, pica2016RPB, bertrand2016NatNano, shang2013APL, zeng2013SR, calderon2017PRL}.  In the larger context of semiconductor nanostructure physics, quantum coherent transport between quantum dots and nanowires could help characterize coherent properties of electronic states, and could have wide ranging applications, such as in the search and control of possible Majorana excitations in hybrid structures \citep{weymann2017PRB,albrecht2017PRL}.

In this paper, we make a thorough examination of quantum coherent electron transport in a silicon double quantum dot (DQD). We first clarify the energy spectrum of the valley-orbital degrees of freedom, then consider the coupled valley and orbital dynamics.  The multiple level anti-crossings provide ample opportunities for Landau-Zener (LZ) transitions and Landau-Zener-Stückelburg (LZS) interference \citep{rubbmark1981PRA, shevchenko2010PhysRep, petta2010Science, studenikin2012PRL, takada2014PRL, bautze2014PRB}.  Accurate measurement of tunnel coupling in a Si double dot, especially the inter-valley tunnel coupling, has great importance to the characterization of the DQD.  Here we extend the widely used DiCarlo method \citep{DiCarlo2004PRL} to a four-level theory for a Si DQD. Besides, with valley-orbit phase difference between two quantum dots of particular importance to electron tunneling and exchange coupling, we propose several schemes to detect this valley phase difference in a Si DQD, ranging from conventional tunneling current measurement to schemes that take advantage of LZ transitions and LZS interferences through multiple level anti-crossings of the DQD.

We then focus on spin transport in a silicon double dot, where spin transfer fidelity is the core concern \citep{nakajima2018NC}.  Specifically, we investigate both phase and spin flip errors in the transport.  For phase error, we calculate the modified electron $g$-factor due to the double dot potential, and find that the resulting phase error could be notable under certain conditions.  We show an example where a small phase correction near the spin-valley anti-crossing could result in a significant coherence loss. Furthermore, due to spin-valley mixing \citep{rohling2012NJP}, there exist situations where quantum information (phase or coherent superposition) can be lost even though classical information (spin population) is faithfully transported. As for spin flip error, we first investigate spin flip caused by SOC \citep{bychkov1984JPC,dresselhaus1955PR}, which is usually slow due to the weak SOC in Si. However, the presence of valleys gives rise to a level anti-crossing that is particularly broad in the interdot detuning, such that considerable spin flip can occur when the double dot is swept through such an anti-crossing.  We also identify four regions in the valley-splitting-Zeeman-splitting parameter space, and examine the anti-crossings and the resultant spin flip caused by an inhomogeneous magnetic field (presumably generated by a micromagnet) in these regions.  Last but not least, we propose a scheme to probabilistically suppress spin flip errors by using Landau-Zener transitions.

The rest of the paper is organized as follows: In Sec.~\ref{sec:model} we describe the double quantum dot model and our protocol for electron transport. In Sec.~\ref{sec:ValleyOrbit} we consider electron charge dynamics in a double dot involving the orbital and valley degrees of freedom.  We first clarify the low-energy spectrum of a single electron in the double dot, and propose several schemes to detect tunnel coupling and the valley phase difference between two dots.
In Sec.~\ref{sec:PhaseError}, we study phase error caused by corrections on the effective $g$-factor from the double dot confinement potential, and show an example of how a phase correction causes significant fidelity loss through spin-valley mixing. In Sec.~\ref{sec:SpinFlipError}, we study spin flip error caused by either SOC or inhomogeneous field and propose a scheme to suppress such errors using LZ transitions. At last, we conclude in Sec.~\ref{sec:Conclusion}.

\section{\label{sec:model}Single-electron dynamics in a silicon double quantum dot}

\begin{figure}
\includegraphics[width=0.85\columnwidth]{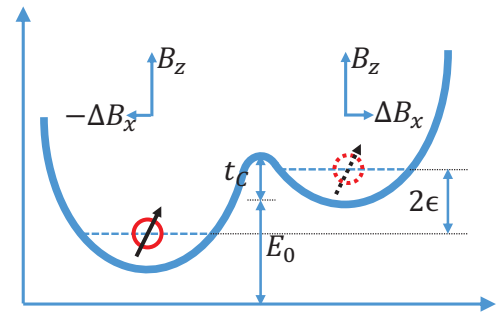}

\caption{\label{fig:scheme}(color online) Scheme of transport in SiQD. By
applying a detuning electric field, the DQD potential minimum can
be tuned as shown in the figure. The detuning between left dot and
right dot is $2\epsilon$, the tunneling barrier is $t_{C}$, and
the ground state energy at
magnetic field $B_{z}$ in $z$-direction and inhomogeneous magnetic
field $\pm\Delta B_{x}$ in $x$-direction.}
\end{figure}

In this work, we consider a simple electron transport protocol by tuning an inter-dot electric field in a Si DQD. The protocol is schematically described in Fig.~\ref{fig:scheme}.  A single electron is confined in a double potential well whose interdot detuning $\epsilon$ is controlled by external gates.  The tunnel barrier $t_{C}$ between the dots can also be controlled by a top gate. The external magnetic field is uniform in the $z$-direction (quantization direction), while a micromagnet provides an inhomogeneous field in the $x$-direction. By changing the interdot detuning $\epsilon$ slowly from $-\epsilon_{0}$ to $+\epsilon_{0}$, the electron, and therefore its spin state, can be adiabatically transported from one dot to the other.

An electron confined in a Si DQD has three degrees of freedom: spin, valley, and orbital.  In the low-energy sector, each involves only two states. The two dots each has a ground orbital state, which we label as the basis $|L\rangle$ and $|R\rangle$. Interface scattering along the growth direction couples the two lowest-energy valley states $|z\rangle$ and $|\bar{z}\rangle$ for the Si conduction band \citep{ando1982RMP} (the other four valleys have higher energy because of the interface confinement, and are not considered in our study here), although the coupling is generally different in different quantum dots. Lastly, the spin of the electron along $z$-direction also has two possible states $|\downarrow\rangle$ and $|\uparrow\rangle$. The tensor of all three degrees of freedom gives us the basis in the single-electron Hilbert space $\{|D,\xi,\sigma\rangle\}$. To avoid information loss by the spin qubit, our transport protocol has to be operated slowly relative to the orbital excitation energy, so that we can limit ourselves to the low-energy sector of the Hilbert space, with the minimal basis set of $D=L,\ R$, $\xi=z,\ \bar{z}$, and $\sigma=\downarrow,\ \uparrow$.  In this basis, the total Hamiltonian of the DQD is expressed as
\begin{widetext}
\begin{equation}
H=E_{0}+\left[\begin{array}{cccccccc}
\epsilon+E_{z} & S_{1}+E_{x} & \Delta_{L} & 0 & t_{C} & S_{2} & 0 & 0\\
S_{1}+E_{x} & \epsilon-E_{z} & 0 & \Delta_{L} & -S_{2} & t_{C} & 0 & 0\\
\Delta_{L}^{*} & 0 & \epsilon+E_{z} & S_{1}+E_{x} & 0 & 0 & t_{C} & S_{2}\\
0 & \Delta_{L}^{*} & S_{1}+E_{x} & \epsilon-E_{z} & 0 & 0 & -S_{2} & t_{C}\\
t_{C} & -S_{2} & 0 & 0 & -\epsilon+E_{z} & -S_{1}-E_{x} & \Delta_{R} & 0\\
S_{2} & t_{C} & 0 & 0 & -S_{1}-E_{x} & -\epsilon-E_{z} & 0 & \Delta_{R}\\
0 & 0 & t_{C} & -S_{2} & \Delta_{R}^{*} & 0 & -\epsilon+E_{z} & -S_{1}-E_{x}\\
0 & 0 & S_{2} & t_{C} & 0 & \Delta_{R}^{*} & -S_{1}-E_{x} & -\epsilon-E_{z}
\end{array}\right]\,.\label{eq:HSVO}
\end{equation}
\end{widetext}
Here $E_{0}$ is the ground state energy for either of the DQD at zero detuning (setting a reference point for energy). The interdot tunnel coupling between the two orbital states $|L\rangle$ and $|R\rangle$ (of the same valley) is labeled as $t_{C}$, and the detuning between the two dots is given by $\epsilon$. The off-diagonal elements $\Delta_{D}=|\Delta_{D}|e^{i\phi_{D}}$ ($D=L,R$) is the valley-orbit coupling connecting the two valleys in each of the two dots with corresponding valley phase $\phi_{D}$.  The Zeeman splitting along $z$ is $E_{z}=\frac{1}{2}g\mu B_{z}$, while the inhomogeneous magnetic field in the $x$-direction leads to a position-dependent splitting of $\pm E_{x}=\pm\frac{1}{2}g\mu B_{x}$, which is used to generate spin rotation through electric dipole spin resonance (EDSR) \citep{benito2017PRB,pioro2008NatPhys,thalineau2014PRB}.
Lastly, the SOC matrix elements are $S_{1}=\langle L,\xi,\uparrow|H_{SO}|L,\xi,\downarrow\rangle$ and $S_{2}=\langle L,\xi,\uparrow| H_{SO} |R,\xi,\downarrow \rangle$, where $H_{SO}$ is the SOC Hamiltonian \citep{bychkov1984JPC,dresselhaus1955PR}.  Here only the matrix elements for orbital ground states $s$ are included.  Coupling to higher energy orbital states will be discussed in Sec.~\ref{subsec:effg} in the context of the effective $g$-factor for the electron spin.

Among the parameters in Hamiltonian (\ref{eq:HVOD}), interdot detuning $\epsilon$, tunnel coupling $t_{C}$, and the uniform Zeeman splitting $E_{z}$ are the most easily tunable experimentally using top gates or the applied magnetic field \citep{noiri2017SST}. The inhomogeneous transverse magnetic field $E_{x}$ can be adjusted by redesigning the micromagnet. It has been shown in Si/SiO$_{2}$ samples that valley-orbit coupling $\Delta_{D}$ and the spin-orbit matrix elements $S_{1}$ and $S_{2}$ can be tuned by the interface electric field \citep{yang2013NatComm, ciftja2016AIPAdv, zimmerman2017NanoLett, goswami2007NatPhys}, though similar results have not been reported for Si/SiGe samples.  In the following discussion, we assume $|\Delta_{D}|$, $\phi_{D}$, $S_{1}$, and $S_{2}$ are fixed for a particular DQD, and keep the rest tunable.

The transport protocol we consider is driven by changing the detuning $\epsilon$ \citep{skinner2003PRL,taylor2005NatPhys,lu2003Nature}.  Specifically, we consider an increasing detuning from $-\epsilon_{0}$ to $\epsilon_{0}$. Initially, the detuning is negative, $|L\rangle$ has lower energy, and the electron is trapped in the left dot. As the detuning adiabatically increases to a positive value, $|R\rangle$ eventually has lower energy and the electron would tunnel to the right dot. The electron evolution during the transport is governed by the time-dependent Schrödinger equation
\begin{equation}
i\hbar\frac{d}{dt}|\psi(t)\rangle=H(t)|\psi(t)\rangle\,.
\end{equation}
In our study, we numerically solve the time-dependent Schrödinger equation for the electron dynamics.  We also diagonalize the Hamiltonian at each time point with detuning $\epsilon(t)$ to obtain the instantaneous eigen-energies $E_{i}(\epsilon)$ and the corresponding eigen-states $|\psi_{i} (\epsilon) \rangle$. These instantaneous eigen-states are helpful in the study of LZ transitions at level anti-crossings, which is a recurring topic through this paper.

\section{\label{sec:ValleyOrbit}Valley-orbit spectrum and dynamics}

In this Section we focus on the charge dynamics in a Si DQD.  We first solve for the low-energy orbital spectrum of the one-electron DQD as a function of interdot detuning. We show that valley phase difference between the DQD is a crucial parameter in determining the electron states and spectrum of a Si DQD, and allows tunnel coupling between any pair of valley eigenstates in the two dots.  we then examine closely how the tunnel couplings between different states can be determined, and propose an extension of the well established DiCarlo method \citep{DiCarlo2004PRL} to the multi-valley situation of a Si DQD.  We also propose several schemes to measure the valley phase difference between two quantum dots based on DC transport, LZ transitions, and LZS interference in a charge-sensing experiment. Our results here should help pave the way toward a quantitative experimental investigation of the valley phase difference.

\subsection{\label{ssec:OrbitSpectrum}Orbital energy spectrum of a Si DQD}

Focusing on the low-energy charge dynamics of a Si DQD, the full Hamiltonian in Eq.~(\ref{eq:HSVO}) can be reduced to a single-electron valley-orbit
Hamiltonian
\begin{equation}
H_{VO}=\left[\begin{array}{cccc}
E_{0}+\epsilon & \Delta_{L} & t_{C} & 0\\
\Delta_{L}^{*} & E_{0}+\epsilon & 0 & t_{C}\\
t_{C} & 0 & E_{0}-\epsilon & \Delta_{R}\\
0 & t_{C} & \Delta_{R}^{*} & E_{0}-\epsilon
\end{array}\right]\,.\label{eq:HVO1}
\end{equation}
In general each single quantum dot in Si has its own complex valley-orbit coupling $\Delta$. With tunnel coupling a perturbation, we first solve the single-dot valley-orbit Hamiltonian, then recast the DQD Hamiltonian (\ref{eq:HVO1}) over the single-dot eigenbasis. Specially, the eigenstates of a single dot Hamiltonian (for instance, the left dot) $H_{L}=\left[\begin{array}{cc}
E_{0}+\epsilon & \Delta_{L}\\
\Delta_{L}^{*} & E_{0}+\epsilon
\end{array}\right]$ are $|L,\pm\rangle\equiv\frac{1}{\sqrt{2}}(|L,z\rangle\pm e^{i\phi_{L}}|L,\bar{z}\rangle)$
\citep{culcer2010PRB2,culcer2010PRB}. Using the set of new basis \{$|L,+\rangle$, $|L,-\rangle$, $|R,+\rangle$, $|R,-\rangle$\},
the valley-orbit Hamiltonian (\ref{eq:HVO1}) now takes the form
\begin{align}
 & H_{VO}^{\prime}=\nonumber \\
 & E_{0}+\left[\begin{array}{cccc}
\epsilon+|\Delta_{L}| & 0 & t_{+} & t_{-}\\
0 & \epsilon-|\Delta_{L}| & t_{-} & t_{+}\\
t_{+}^{*} & t_{-}^{*} & -\epsilon+|\Delta_{R}| & 0\\
t_{-}^{*} & t_{+}^{*} & 0 & -\epsilon-|\Delta_{R}|
\end{array}\right]\,,\label{eq:HVOD}
\end{align}
where $t_{+}=\frac{1}{2}t_{C}[1+e^{-i(\phi_{L}-\phi_{R})}]$ and $t_{-}=\frac{1}{2}t_{C}[1-e^{-i(\phi_{L}-\phi_{R})}]$ are the intra- and inter-valley tunnel couplings, respectively. Here by ``valley'' we mean the single-dot valley eigenstates instead of the $z$ and $\bar{z}$ bulk valley states. Clearly, both tunnel couplings $t_{-}$ and $t_{+}$ are sensitively dependent on the valley phase difference $\delta\phi\equiv\phi_{L}-\phi_{R}$ between the two dots, and they in turn determine the eigenstates and energies of the DQD.  Tunnel couplings $t_{\pm}$ here are complex numbers, while
the real couplings are also widely used \citep{burkard2016PRB}. It can be proven that Hamiltonian (\ref{eq:HVOD}) can be written in the form used in Ref.~\citep{burkard2016PRB} by applying a rotation of the reference frame.

\begin{figure}
\includegraphics[width=1\linewidth]{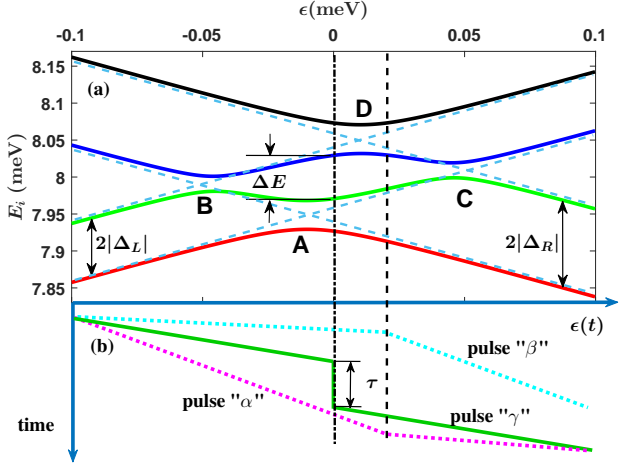}

\caption{\label{fig:EngDiagram}(color online) (a) Energy levels as a function
of detuning $\epsilon$. (b) Applied pulses of $\epsilon$ as functions
of time. The parameters are chosen as $|\Delta_{L}|=40\,\mbox{\ensuremath{\mu}eV}$,
$|\Delta_{R}|=60\,\mbox{\ensuremath{\mu}eV}$, $E_{0}=8\,\mbox{meV}$,
$|t_{-}|=40\,\mbox{\ensuremath{\mu}eV}$ $\delta\phi=0.3\pi$.}
\end{figure}

In Fig.~\ref{fig:EngDiagram}~(a) we plot the DQD energy levels as a function of the interdot detuning. Different magnitudes of valley splitting between the two dots make the DQD spectrum here asymmetric across the zero detuning, while the generally non-vanishing inter- and intra-valley tunnelings $t_{-}$ and $t_{+}$ produce four anti-crossings, labeled as ``A'', ``B'', ``C'', and ``D''. If we sweep the interdot detuning (as in our electron transport protocol) through any of these anti-crossings, the probability of a diabatic versus an adiabatic transition is determined by the tunnel coupling $t_{-}$ or $t_{+}$ as compared to the sweeping speed.

Valley-orbit coupling in a Si DQD is essential not only in determining
the electron charge dynamics, but also for spin-based quantum information
processing in Si quantum dots \citep{friesen2007PRB,friesen2006APL,wu2012PRB,boross2016PRB},
and valley phase is of particular importance there, too \citep{zimmerman2017NanoLett}.
One example is discussed later in this manuscript in Sec.~\ref{subsec:SV-Entanglement}.
Therefore, in the current study of charge dynamics in a Si DQD, one
of our major goals is to identify possible ways to measure tunneling
coupling $t_{\pm}$ and valley phase difference $\delta\phi$.

\subsection{\label{subsec:tunneling energy} Determining tunneling couplings
in a Si double dot}

Tunneling energy is a crucial property of a DQD. In a GaAs DQD there
is a well established procedure for measuring tunneling energy through
charge sensing \citep{DiCarlo2004PRL}, with the low-energy orbital
dynamics of a single electron in the DQD described as a two-level
system. Such a procedure has also been applied to Si DQDs to find
the tunnel coupling between the lowest-energy orbital states in the
two dots \citep{simmons2009NanoLett}. However, in a Si DQD, the low-energy
dynamics of the DQD should in general be described by a four-level
model as in Eq.~(\ref{eq:HVOD}) to include the valley information,
since valley splitting is usually much smaller than single-dot intra-valley
orbital excitation energy. Furthermore, DiCarlo's formula can only be used to measure intra-valley
$|t_{+}|$. To measure the inter-valley $|t_{-}|$ and to increase
measurement accuracy, it is necessary to develop a new theory that
account for all tunneling energies in a Si DQD.

Within DiCarlo's method, tunnel coupling is obtained by curve fitting
from the charge distribution measurements as inter-dot detuning is
varied. We adopt the same approach but base our theory on four relevant
energy levels instead of two. While direct diagonalization of Hamiltonian~(\ref{eq:HVOD})
is straightforward, it only yields complicated expressions for the
eigenstates that lack intuition, and are cumbersome to use for curve
fitting. Instead, here we derive a natural extension to DiCarlo's
formula that is simple to use and understand.

We set our starting point with an electron in a DQD that has no inter-valley
coupling, i.e. $|t_{-}|=0$ in Eq.~(\ref{eq:HVOD}). In this limit
the four eigen-energies are at $E_{1,\pm}=\pm\Delta_{+}-E_{\pm}$
and $E_{2,\pm}=\pm\Delta_{+}+E_{\pm}$, where $E_{\pm}=\sqrt{(\epsilon\pm\Delta_{-})^{2}+|t_{+}|^{2}}$
and $\Delta_{\pm}=\frac{1}{2}(|\Delta_{L}|\pm|\Delta_{R}|)$. The
corresponding eigen-states are $|\psi_{1,\mp}\rangle=\cos\frac{\theta_{\mp}}{2}|R,\mp\rangle-\sin\frac{\theta_{\mp}}{2}|L,\mp\rangle$
and $|\psi_{2,\mp}\rangle=\sin\frac{\theta_{\mp}}{2}|R,\mp\rangle+\cos\frac{\theta_{\mp}}{2}|L,\mp\rangle$,
where $\tan\theta_{\mp}=\frac{|t_{+}|}{\epsilon\mp\Delta_{-}}$ ($\theta_{\mp}\in[0,\pi]$).
When such a system is cooled down to the ground state $|\psi_{1,-}\rangle$,
the probability of finding the electron in the left dot is
\begin{align}
P_{L}^{(1)} & =\sin^{2}\frac{\theta_{-}}{2}=\frac{1}{2}\left(1-\frac{\epsilon-\Delta-}{2E_{+}}\right)\,,\label{eq:PL1}
\end{align}
which is identical to the expression of charge occupation given in
Ref.~\citep{DiCarlo2004PRL}. Physically, in the absence of inter-valley
tunneling, the two valley eigenstates evolve within their own subspace,
so that the charge distribution is reduced to the two-level case in
GaAs. In the analysis here we have assumed zero temperature for the
electron. Finite temperatures can be straightforwardly accounted for
by adding a thermal broadening factor of $\tanh\left(\frac{E_{+}}{2k_{B}T}\right)$~\citep{DiCarlo2004PRL}.

\begin{figure}
	\includegraphics[width=1\linewidth]{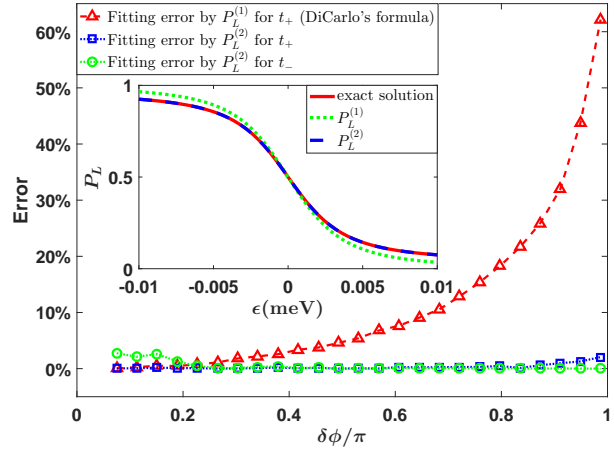}
	
	\caption{\label{fig:FitError}(color online) Curve fitting error by using two
		different schemes. The parameters are chosen as $|\Delta_{L}|=100\,\mbox{\ensuremath{\mu}eV}$,
		$|\Delta_{R}|=100\,\mbox{\ensuremath{\mu}eV}$, $t_{C}=50\,\mu{\rm eV}$
		for main plot. The inset figure shows the accuracy of $P_{L}^{(1)}$
		and $P_{L}^{(2)}$, with a set of particular parameters $|t_{+}|=7.8\,\mu{\rm eV}$
		and $|t_{-}|=49.4\,\mu{\rm eV}$.}
\end{figure}

When the inter-valley tunneling $|t_{-}|$ is finite, the Hamiltonian
$H_{VO}^{\prime}$ can be diagonalized in the basis $\{|\psi_{1,\mp}\rangle,|\psi_{2,\mp}\rangle\}$.
The approximate ground state $|g\rangle\approx\cos\frac{\Theta}{2}|\psi_{1,-}\rangle-\sin\frac{\Theta}{2}|\psi_{2,+}\rangle$
($\tan\Theta=\frac{|t_{-}|}{E_{+}+E_{-}+\Delta_{+}}$) can be obtained
by neglecting the matrix elements in $H_{VO}^{\prime}$ with the factor
$\sin(\frac{\theta_{+}+\theta_{-}}{2})$. The probability of finding
the electron in the left dot now becomes
\begin{equation}
P_{L}^{(2)}=\cos^{2}\frac{\Theta}{2}\sin^{2}\frac{\theta_{-}}{2}+\sin^{2}\frac{\Theta}{2}\cos^{2}\frac{\theta_{+}}{2}\,.\label{eq:PL2}
\end{equation}
In the inset of Fig.~\ref{fig:FitError}, we compare the accuracy
of $P_{L}^{(1)}$ and $P_{L}^{(2)}$ to the exact solution (full diagonalization
of the $H_{VO}^{\prime}$). It shows that the approximation $P_{L}^{(2)}$
has a much better agreement with the exact solution than $P_{L}^{(1)}$.

In the DiCarlo approach the tunnel coupling is obtained via curve
fitting from charge distribution measurement using Eq.~(\ref{eq:PL1}).
Here for a Si DQD there are two tunnel coupling energies $|t_{\pm}|$.
They can again be obtained through curve fitting over a charge measurement,
such as the solid red $P_{L}$ curve in the inset of Fig.~\ref{fig:FitError}.
If $t_{+}$ is the only quantity of interest, one can attempt to use
both Eq.~(\ref{eq:PL1}) and Eq.~(\ref{eq:PL2}) to fit the solid
red curve with fitting parameters $|t_{\pm}|$.
Assuming the best fit is obtained with $|t_{\pm}^{best}|$, we can
evaluate the accuracy of Eq.~(\ref{eq:PL1}) and Eq.~(\ref{eq:PL2})
using percentage error $\frac{\left||t_{\pm}^{best}|-|t_{\pm}|\right|}{|t_{\pm}|}$.
As shown in Fig.~\ref{fig:FitError}, and as we have discussed above,
$P_{L}^{(1)}$ (DiCarlo's formula) is only valid for $\delta\phi\approx0$,
when $|t_{-}|\approx0$. If $\delta\phi\neq0$, particularly when
$\delta\phi\rightarrow\pi$, the fitting error for $|t_{+}|$ with
$P_{L}^{(1)}$ in Eq.~(\ref{eq:PL1}) is significant. In other words
it is generally necessary to employ $P_{L}^{(2)}$ in Eq.~(\ref{eq:PL2})
in order to obtain $t_{+}$ accurately. Furthermore, only by using
$P_{L}^{(2)}$ can one obtain $|t_{-}|$, as it does not appear in
$P_{L}^{(1)}$.

Mathematically, $P_{L}^{(1)}$ is obtained by setting $|t_{-}|=0$.
Physically, it can be interpreted as neglecting the impact of anti-crossings
``B'' and ``C'' (see Fig.~\ref{fig:EngDiagram}) on the charge
distribution. On the other hand, $P_{L}^{(2)}$ does incorporate corrections
from those two anti-crossings, and produces better fittings inevitably.

\subsection{Valley phase difference in a Si double dot}

As we point out in Sec.~\ref{ssec:OrbitSpectrum}, the difference between the phase of the valley-orbit coupling in each of the double dot determines
the tunnel couplings between different valley eigenstates.  Below we discuss a few different approaches to measure this phase difference.

\subsubsection{Measuring valley phase difference from tunnel couplings $|t_{\pm}|$}

A simple calculation based on Hamiltonian (\ref{eq:HVOD}) shows that
\begin{equation}
\frac{|t_{-}|}{|t_{+}|}=\tan\frac{\delta\phi}{2}.\label{eq:EPEM}
\end{equation}
Therefore, the valley phase difference $\delta\phi$ can be obtained through measuring $|t_{\pm}|$, which can be done through a static charge sensing measurement as we discussed in the previous subsection~\ref{subsec:tunneling energy}.

As shown in a recent experiment \citep{mi2017PRL}, the intra- and inter-valley tunneling rates can also be extracted from the cavity response when a Si DQD with one electron is strongly coupled to a cavity.  The valley phase difference $\delta\phi$ can then be obtained according to Eq.~(\ref{eq:EPEM}) from the measured ratio $\frac{|t_{-}|}{|t_{+}|}$.

Another approach to determine the two tunnel couplings $|t_{-}|$ and $|t_{+}|$ in a Si DQD is via a DC transport experiment, where the DQD is coupled to a source and a drain lead. The resonant tunneling current at the anti-crossings ``A'' and ``B'' (assuming source-dot and dot-drain tunnel couplings are stronger than the interdot tunnel coupling) should then give a good estimate of the interdot tunnel couplings, and therefore the valley phase difference between the two dots~\citep{stoof1996PRB}:
\begin{equation}
\frac{I_{A}}{I_{B}}=\frac{|t_{-}|^{2}}{|t_{+}|^{2}}=\tan^{2}\frac{\delta\phi}{2}\,.
\end{equation}
Alternatively, if the DQD couplings to the source and drain leads are weaker than the interdot coupling, the DQD would act as a single entity.  Tunnel current through the DQD should then display two peaks for the anti-crossing levels, and the peak spacing would then give a measure of the tunnel coupling strength $|t_+|$ or $|t_-|$.  When both tunnel couplings are measured, the valley phase difference can again be obtained.

The setup here is that of a conventional DC transport experiment with electrons tunneling through the DQD sequentially. It does require that the DQD couples to a source and a drain lead. For an electron spin or charge qubit this may not be the optimal arrangement unless the coupling to the leads can be cut off almost completely.

\subsubsection{Measuring valley phase difference $\delta\phi$ by charge sensing}

For a closed double dot without close-by leads, one can obtain information on $|t_{-}|$ and $|t_{+}|$ from LZ transitions across the energy-level anti-crossings. When sweeping the interdot detuning through an anti-crossing, the probability for the electron to follow a diabatic path can be roughly predicted by the general LZ formula~\citep{rubbmark1981PRA},
\begin{equation}
P_{D}=\exp\left(-\frac{2\pi a^{2}/\hbar}{d|E_{m}-E_{n}|/dt}\right)\,,\label{eq:LZF}
\end{equation}
where $a$ is the off-diagonal element coupling the two involved energy levels $E_{m}$ and $E_{n}$, which is also the half energy gap at the anti-crossing. In a transport experiment, the output state is mainly determined by the LZ velocity $v_{LZ}=\frac{\partial}{\partial t}(E_{m}-E_{n})$, which can be controlled by the detuning pulse.

Consider a pulse ``$\alpha$'' depicted by the magenta-dotted line in Fig.~\ref{fig:EngDiagram}~(b), where the interdot detuning is swept through anti-crossing ``A'' relatively slowly, then through anti-crossing ``C'' quickly. Here, ``slowly'' or ``quickly'' are defined by whether the LZ velocity is comparable to the the gap $|t_{\pm}|/2$. After pulse ``$\alpha$'', the total probability of finding the electron in the left dot at large positive detuning (blue line, state $|L\rangle$) is $P_{L\alpha} = \exp \left( -\frac{\pi|t_{-}|^{2}}{2\hbar v_{LZ}^{\alpha}} \right)$, which can be monitored by a charge sensor. If we revert the pulse sequence and design it like the cyan-dotted line in Fig.~\ref{fig:EngDiagram}~(b), labeled as pulse ``$\beta$'' (sweep through ``A'' quickly and ``C'' slowly), the final charge distribution after the pulse sequence will be $P_{L\beta} = \exp \left( -\frac{\pi|t_{+}|^{2}}{2\hbar v_{LZ}^{\beta}} \right)$.

We can choose the same LZ velocities for the front and back half of the $\alpha$ and $\beta$ pulse sequences, respectively, as shown in Fig.~\ref{fig:EngDiagram}~(b), so that $v_{LZ}^{\alpha}=v_{LZ}^{\beta}=v_{LZ}$.  The ratio of logarithms of charge distribution for the two pulses
is then
\begin{equation}
\frac{\ln P_{L\alpha}}{\ln P_{L\beta}}=\frac{|t_{-}|^{2}}{|t_{+}|^{2}}=\tan^{2}\left(\frac{\delta\phi}{2}\right)\,.
\end{equation}
The ratio now is directly related to the valley phase difference, and is independent of both $t_{C}$ and $v_{LZ}$. One can thus perform multiple experiments with different combinations of $t_{C}$ (by tuning the tunnel barrier between the dots) and $v_{LZ}$ to improve the accuracy of this estimate.

The accuracy of $\delta\phi$ obtained with this approach relies on high-precision charge distribution measurement and precise control of the speed of detuning sweep, and is further limited by factors such as orbital relaxation.  In the next subsection, we show that $\delta\phi$ can also be extracted from LZS interference patterns even when the contrast in the interference is limited by relaxation and incomplete initialization.

\subsubsection{Measuring valley phase difference $\delta\phi$ by Landau-Zener-Stückelburg
interference}

The multiple valley-induced level anti-crossings in a Si DQD provide a rich ground for creating and observing LZS interference.  An interference pattern, together with controlled variables such as the detuning sweeping speed, allows the possibility of determining the intra- and inter-valley tunnel couplings, which in turn allow us to calculate the valley phase difference $\delta\phi$ in the DQD. Here we choose one particular interference pattern to measure the energy gap $\Delta E$ at zero detuning (shown in Fig.~\ref{fig:EngDiagram}), and extract information on tunnel coupling and valley phase difference through curve fitting.  The ``zero detuning'' here can be determined by shifting the detuning by the amount of $\frac{1}{2} (|\Delta_{L}| - |\Delta_{R}|)$ from the anti-crossing ``A'', or using the mean value $\frac{1}{2}(\epsilon_{B} + \epsilon_{C})$ for detuning. The eigen-energies of Hamiltonian (\ref{eq:HVO1}) and (\ref{eq:HVOD}) at zero detuning can be analytically calculated from the four-level eigenenergies:
\begin{eqnarray}
& & E_{1} = E_{0}-\sqrt{E_{A}^{2}+E_{B}^{2}},\quad E_{2}=E_{0}-\sqrt{E_{A}^{2}-E_{B}^{2}},\label{eq:EigE12} \\
& & E_{3} = E_{0}+\sqrt{E_{A}^{2}-E_{B}^{2}},\quad E_{4}=E_{0}+\sqrt{E_{A}^{2}+E_{B}^{2}},\label{eq:EigE34}
\end{eqnarray}
where
\begin{widetext}
\begin{eqnarray}
E_{A}^{2} & = & \frac{|\Delta_{L}|^{2}+|\Delta_{R}|^{2}}{2}+t_{C}^{2}\,,\\
E_{B}^{2} & = & \sqrt{\frac{(|\Delta_{L}|^{2}-|\Delta_{R}|^{2})^{2}}{4}+t_{C}^{2}\left[|\Delta_{L}|^{2} +|\Delta_{R}|^{2}+2|\Delta_{L}||\Delta_{R}|\cos(\delta\phi)\right]}\,.
\end{eqnarray}
The relation between the zero-detuning gap $\Delta E=(E_{3}-E_{2})|_{\epsilon=0}$ and valley phase difference $\delta\phi$ can then be obtained as
\begin{equation}
\frac{\Delta E^{2}}{4}=\frac{|\Delta_{L}|^{2}+|\Delta_{R}|^{2}}{2}+\left(\frac{|t_{-}|}{\cos\frac{\delta\phi}{2}}\right)^{2} -\sqrt{\frac{(|\Delta_{L}|^{2}-|\Delta_{R}|^{2})^{2}}{4}+\left(\frac{|t_{-}|}{\cos\frac{\delta\phi}{2}}\right)^{2} \left[\left(|\Delta_{L}|-|\Delta_{R}|\right)^{2}+4|\Delta_{L}||\Delta_{R}|\cos^{2}\frac{\delta\phi}{2}\right]}.\label{eq:DE_dp}
\end{equation}
\end{widetext}
$\delta\phi$ can thus be extracted from measurement of $\Delta E$ if $|\Delta_{L,R}|$ and $|t_{-}|$ can be measured in advance. Specifically, as shown in Fig.~\ref{fig:EngDiagram}, two adjacent LZ processes ``B'' and ``C'' can form an LZS interferometer. Our designed pulse sequence ``$\gamma$'' has a plateau at $\epsilon=0$ as shown in Fig.~\ref{fig:EngDiagram}~(b), which leads to tuning of the charge distribution of the output state. In order for the electron to pass through both ``B'' and ``C'' anti-crossings, the system needs to be at least partially prepared in the second lowest energy state $|2\rangle\approx|L,+\rangle$ (green on the left side) at $\epsilon\ll-t_{C}$. After the pulse sequence, the probabilities of finding $|L,-\rangle$ (blue on the right side) and $|R,+\rangle$ (green on the right side) in the final state is then strongly dependent on the dynamical phase $\exp(-i\Delta E \tau/\hbar)$ accumulated at $\epsilon=0$, which can be tuned by $\tau$ and monitored by a charge sensor \citep{simmons2009NanoLett, petta2004PRL}.

\begin{figure}
\includegraphics[width=1\linewidth]{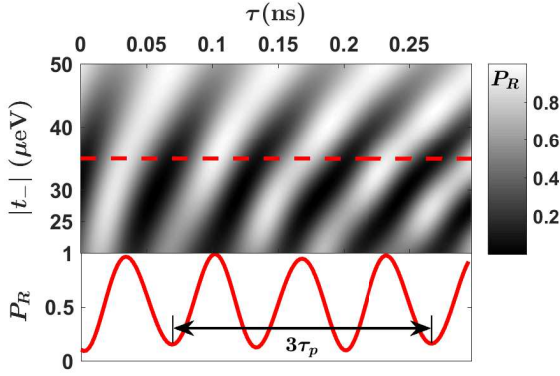}

\caption{\label{fig:Interference}(color online) Probability of finding the
electron in the right dot after the LZS process ($P_{R}=|\langle R|\psi_{final}\rangle|^{2}$)
as a function of the tunneling energy $t_{-}$ and the waiting time
$\tau$. The lower panel is a cross-section of the 3-d contour plot
at $|t_{-}|=35\ {\rm \mu eV}$. The parameters are chosen as $|\Delta_{L}|=60\,\mbox{\ensuremath{\mu}eV}$,
$|\Delta_{L}|=40\,\mbox{\ensuremath{\mu}eV}$, $\delta\phi=0.3\pi$.
$\epsilon(t)$ is changing from $-0.2$ meV to $0.2$ meV in an operation
time (excluding $\tau$) $T-\tau=1$ ns.}
\end{figure}

In Fig.~\ref{fig:Interference} we plot the probability of finding the electron in the right dot $P_{R}$ in the final state as a function of the tunnel coupling between the lower-energy valley eigenstates $t_{-}$ and pulse plateau duration $\tau$. As expected, the figure clearly shows the interference between the two LZ transitions at ``B'' and ``C''. The period $\tau_{p}$ as shown in Fig.~\ref{fig:Interference} corresponds to a total accumulated phase $2\pi$. The energy gap $\Delta E$ at $\epsilon=0$ is then obtained as $\Delta E=E_{3}-E_{2}|_{\epsilon=0}=2\hbar\pi/\tau_{p}$.  It is worth noting that a similar scheme using interference pattern to measure energy gap has been demonstrated in a recent experiment \citep{schoenfield2017NatComm}.

The LZS interference discussed here provides a robust approach to measure the energy difference between energy levels 2 and 3.  In general one could use it to build an accurate map of $E_3(\epsilon) - E_2(\epsilon)$.  On the other hand, in order to determine $\delta\phi$, the measurement of $\Delta E$ here has to be combined with other experiments that measures tunnel coupling $|t_{-}|$ and single-dot valley splittings $|\Delta_{L}|$ and $|\Delta_{R}|$.  From this perspective it cannot act as a stand-alone measurement of $\delta \phi$, but can work as a good verification tool.
Indeed, the intra-valley tunneling coupling $t_{-}$ is part of the charge sensing measurement we have discussed above \citep{petta2004PRL, DiCarlo2004PRL, simmons2009NanoLett}.  With knowledge of $|\Delta_{L}|$, $|\Delta_{R}|$, and $|t_{-}|$, one can then determine $\delta\phi$ by either solving Eq.~(\ref{eq:DE_dp}) or directly fitting the original experimental data of $\Delta E$ and $t_{-}$, and the result can be directly compared to the charge sensing fitting results to assure their reliability.

The robustness of the LZS interference measurement lies in the fact that it is a pattern measurement (changes in $P_{R}$) instead of an intensity (such as $P_{R}$ itself) measurement.  For example, the anti-crossing ``A'' in Fig.~\ref{fig:EngDiagram} could divert the electron into an irrelevant state (the ground state) as the detuning sweeps across it. However, such a probability leakage only lowers the contrast of the interference pattern, but does not changes its period.  Similarly, an incomplete preparation into state 2 initially also only leads to a reduction in the contrast of the interference pattern.  As such the LZS interference approach does have its advantage in determining accurately the interdot tunnel couplings and the valley phase difference.

A key ingredient of the LZS interference measurement is the accumulated phase, which is sensitively dependent on the energy gap and can be affected by the charge noise \citep{freeman2016APL,yoneda2018NatNano}.  However, at zero detuning where the extra phase is accumulated in our protocol, the first order derivative of the energy gap with respect to detuning is zero, namely $\frac{d\Delta E}{d\epsilon}=0$ at $\epsilon=0$.  This can be verified either from Eq.~(\ref{eq:EigE12}-\ref{eq:EigE34}) or from Fig.~\ref{fig:EngDiagram}~(a). Therefore, the phase delay in our proposal is robust against charge noise in the leading order.

In short, in this section we have studied the electron charge spectrum and dynamics in a Si DQD, accounting for both valley and orbital degrees of freedom. We have discussed several possible schemes to measure the tunneling energies as well as valley phase difference of a Si DQD. These proposals should open new paths toward understanding of the valley properties of Si DQD samples.

\section{Spin transport in a double dot: Phase errors \label{sec:PhaseError}}

When transporting an electron spin qubit from one dot to another, it is crucial to maintain the spin state in which quantum information is encoded while shifting the electron's location. For example, errors could arise in the accumulated dynamical phase during the transport due to variations in spin splitting.  In addition, the presence of valleys in Si means additional anti-crossings among the energy levels as interdot detuning is changed, which could lead to spin-valley mixing and further coherence loss.

Here we examine two important phase errors that can occur during spin transport in a Si DQD.  The first is caused by the variations in the effective $g$-factor of the electron.  When a superposed spin state is transported in a Si DQD, the accumulated phase between the two spin orientations is determined by the confinement-potential-dependent $g$-factor. We show that if this variation in the $g$-factor is not properly accounted for, the error in the accumulated phase can be notable under certain conditions.  The second phase-related issue we investigate is the coherence loss due to spin-valley mixing, where a small phase difference between valleys (due to valley-dependent $g$-factor) can cause significant fidelity loss through spin-valley anti-crossings.

\subsection{\label{subsec:effg}DQD Corrections on the effective $g$-factor}

To transfer a spin qubit with high fidelity, one requirement is to keep the orbital degree of freedom frozen. This is typically achieved by sweeping the interdot detuning slowly and keeping the electron orbital dynamics adiabatic. Assuming an absence of nuclear spins and other magnetic defects, and keeping in mind that the electron spin $g$-factor is dependent on the potential it experiences, the phase of the excited spin state is $\int_{0}^{t} g_{eff}(\tau) \mu_{B}B d\tau$.  In other words, the accumulated dynamical phase is dependent on the adiabatic path for the electron in moving from one dot to the other.  Here the effective $g$-factor \citep{buonacorsi2018QST,jock2018NatCommun} is defined as
\begin{equation}
g_{eff}=(E_{g,-,\uparrow}-E_{g,-,\downarrow})/\mu_{B}B\,,
\end{equation}
where $E_{g,-,\uparrow}$ and $E_{g,-,\downarrow}$ are the energies for spin up and down states respectively, while the orbital state is the ground orbital and valley states.

To clarify the degree of modification to the spin qubit dynamical phase, here we calculate the corrections to the electron $g$-factor by the double dot confinement potential. We extend the Hilbert space for Hamiltonian~(\ref{eq:HSVO}) to include higher orbital states in order to calculate the effective $g$-factor, with the details of the calculation shown in Appendix~\ref{sec:Calculation-geff}.  In Fig.~\ref{fig:geff}, we plot the correction on the effective $g$-factor $\delta g=g_{eff}-g_{s}$, with the single-dot $g_{s}$ as a benchmark. Panel (a) shows $\delta g$ as a function of the applied magnetic field. The most prominent features here are the two discontinuities at $B_{c1}$ and $B_{c2}$, where the electron spectrum has an anticrossing due to spin-valley and spin-orbit interaction, respectively. At these anti-crossings $g$-factor is ill-defined, with spin states completely mixed with either valley or orbital states. Away from the anti-crossings we define the spin of a state as its dominant spin direction, which leads to the $g$-factor jumps when the dominant spin direction changes in the states when the interdot detuning is shifted past the anti-crossings. At large detunings (either positive or negative), the electron is strongly confined in one of the quantum dots, so that $\delta g \rightarrow 0$. At zero detuning, the double dot potential produces the largest correction on the effective $g$-factor as expected.  The numerical value of $\delta g$ depends mainly on the confinement potential, but also on other parameters such as valley splitting and the applied magnetic field if spin-valley coupling is sufficiently strong.

\begin{figure}
\includegraphics[width=1\linewidth]{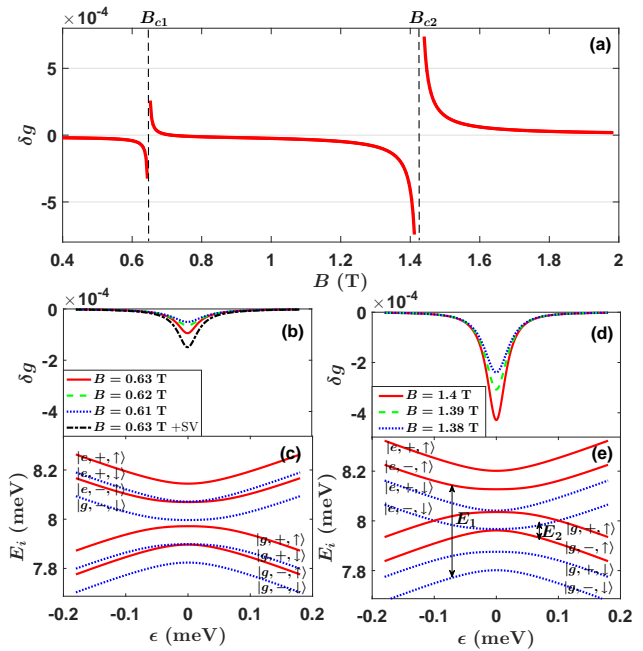}

\caption{\label{fig:geff}(color online) Features of the corrections on effective $g$-factor. (a) $\delta g$ at different $B$-field (plotted at zero
detuning $\epsilon=0$.  Panels (b) \& (d) show $\delta g$ as a function of the
detuning $\epsilon$, while panels (c) \& (e) show energy diagrams near $B_{c1}$and
$B_{c2}$.  The blue lines in (c) and (e) represent spin down states while the red lines represent spin up states. The DQD parameters are chosen
as $l_{0}=10$~nm, $d=20\,{\rm nm}$, $\alpha_{BR}=0.05\,{\rm meV\cdot\mathring{A}}$,
$\alpha_{D}=0.45\,{\rm meV\cdot\mathring{A}}$, valley splitting $|\Delta|=50\,{\rm \mu eV}$,
and the phase difference $\delta\phi=0.4\pi$.}
\end{figure}

When the detuning is uniformly increased from $-\epsilon_{0}$ to $\epsilon_{0}$ , i.e., $\epsilon(\tau)=\epsilon_{0}\frac{\tau}{T}$ ($\tau\in[-T,T]$), the corresponding phase correction accumulated is
\begin{equation}
\Phi_{error}=\int_{-T}^{T}\frac{1}{\hbar}\left[g_{eff}(\tau)-g_{s}\right]\mu_{B}Bd\tau,
\end{equation}
For a typical set of parameters \citep{raith2011PRB,yang2013NatComm} shown in Fig.~\ref{fig:geff}, the correction is calculated in the table~\ref{tab:PhaseError} for a particular B-field $B=1.4\;{\rm T}$.  It indicates that if $g_s$ is used to estimate the dynamical phase, a notable error would result. When estimating the accumulated dynamical phase, the correction from the double dot potential must be taken into consideration.

\begin{table}[H]
\begin{centering}
\begin{tabular}{|c|c|c|c|}
\hline
$\epsilon_{0}$ (meV)  & 0.18  & 0.12  & 0.06\tabularnewline
\hline
\hline
$\Phi_{error}$  & $-0.046\pi$  & $-0.068\pi$  & $-0.125\pi$\tabularnewline
\hline
\end{tabular}
\par\end{centering}
\caption{\label{tab:PhaseError}Phase correction due to the double dot potential for different detuning range in the transport.}
\end{table}

There are several interesting features in the numerical results shown in Fig.~\ref{fig:geff}, which can be explained by the second order perturbation theory,
\begin{equation}
E_{i}=\epsilon_{i}+\langle i|H_{SO}|i\rangle+\sum_{j\neq i}\frac{|\langle i|H_{SO}|j\rangle|^{2}}{\epsilon_{i}-\epsilon_{j}}.\label{eq:Perturbation}
\end{equation}
The first order perturbation $\langle i|H_{SO}|i\rangle$ is always zero, so that the lowest-order correction comes from the second-order term $\sum_{j\neq i} \frac{|\langle i|H_{SO}|j\rangle|^{2}}{\epsilon_{i}-\epsilon_{j}}$.  For example, near $B=B_{c2}\approx1.42\;{\rm T}$ the major correction comes from the SOC, thus the spin splitting can be written as
\begin{equation}
\delta E_{g,-,\uparrow}-\delta E_{g,-,\downarrow}=\frac{|\langle g,\uparrow|H_{SO}|e,\downarrow\rangle|^{2}}{E_{g,-,\uparrow}-E_{e,-,\downarrow}}+\frac{|\langle g,\downarrow|H_{SO}|e,\uparrow\rangle|^{2}}{E_{g,-,\downarrow}-E_{e,-,\uparrow}},\label{eq:gap}
\end{equation}
where $|g\rangle$ and $|e\rangle$ represent the ground and first excited orbital states.  At zero detuning, they are approximately $|g\rangle \approx \frac{1}{\sqrt{2}} (|L\rangle+|R\rangle)$ and $|e\rangle\approx\frac{1}{\sqrt{2}}(|L\rangle-|R\rangle)$.  For a DQD, the major contributions to the energy shifts come from excited states with gaps $E_{1} = E_{e,-,\uparrow} - E_{g,-,\downarrow}$ and $E_{2} = E_{e,-,\downarrow} - E_{g,-,\uparrow}$, are shown in Fig.~\ref{fig:geff}~(e).

With the ground-excited energy gap at the minimum when the double dot detuning is at zero, the largest $|\delta g|$ always occurs at zero detuning,
as shown in Fig.~\ref{fig:geff}~(b) and (d).

Notice that $\delta g$ here is calculated relative to the single-dot $g_s$. In a single dot, the first excited orbital state is the ``$p$'' orbital with an energy gap in the order of the orbital excitation energy $\hbar\omega_{0}$ (See Appendix \ref{sec:Calculation-geff}).
In a double dot, on the other hand, the gap is in the order of tunneling energy $t_{C}$ (assuming the Zeeman splitting is much smaller),
and in general $t_C \ll \hbar \omega_0$.  This is the main reason why a DQD potential causes a correction to the effective $g$-factor.

The correction on $g$-factor sharply increases when the $B$-field approaches certain values, when the Zeeman splitting matches the valley splitting or the tunnel splitting. Theses anti-crossings have been shown theoretically and experimentally to lead to large relaxation rate called ``hot spots'' \citep{raith2011PRB,yang2013NatComm,Zhao2018SR} because of the complete spin-orbit or spin-valley mixing.

Notice that $\delta g$ is finite and not diverging at the two critical $B$-fields. When the gap between two energy levels vanishes, the non-degenerate perturbation expression of Eq.~(\ref{eq:Perturbation}) needs to be replaced by a degenerate perturbation calculation. Furthermore, at the critical $B$-fields, the two spin states are fully mixed so that one cannot define a spin orientation for each state.  Consequently $g$-factor is not well defined at those two fields.

The correction near $B_{c1}$ is smaller than $B_{c2}$ because the first ``hot spot'' is caused by the spin-valley mixing that is a result of different valley mixings in the different quantum dots.  It is limited by the tunnel coupling strength and is much weaker than the intra-valley spin-orbit coupling.

In Refs.~\citep{yang2013NatComm,Huang2014PRB}, the authors have also revealed the existence of a direct SV coupling and the resulting hot spots. The strength of this direct SV coupling is typically in the order of tens of neV, which is about one order of magnitude smaller than the intra-valley SOC (hundreds of neV). In Fig.~\ref{fig:geff}~(b), we plot an additional dash-dotted black curve by taking this direct SV coupling (at a magnitude of $50$ neV) into account. In general, the effect of the direct SV coupling \citep{yang2013NatComm,Huang2014PRB} is negligible except near the hot spot, where it does lead to a notable change to the correction on the effective $g$-factor.

The correction $\delta g$ is positive in some regions while negative in others, with the most dramatic switch happening at the critical fields.  This is a result of the level anti-crossings at the those fields.
When $B_{c1}<B<B_{c2}$, the total correction is a combination of the SV and SO corrections, so that it gradually changes from positive (near $B_{c1}$ with $B > B_{c1}$) to negative (near $B_{c2}$ with $B < B_{c2}$). Since SO correction is stronger as we have discussed above, the total correction is mostly negative in this region.

The SOC strength in our study is chosen conservatively \citep{raith2011PRB}.  Recent experimental studies observe much larger SOC strength in a Si heterostructure near the Si-barrier interface \citep{jock2018NatCommun}, which implies that the corrections on $g$-factor could be notable in a wider range of parameters. In addition, our study here focuses on the impact of SOC on the effective $g$-factor.  If an inhomogeneous magnetic field $E_{x}$ exists, it may cause an even larger correction since its strength is typically one or two order large than SOC \citep{hu2012PRB}, not to mention the usual presence of gradient in $E_z$, which causes further change in the overall Zeeman splitting.  All these factors need to be taken into consideration when estimating dynamical phase change in a transport process.

In summary, we have discussed the correction on the effective $g$-factor and the resulting phase error due to SOC and the DQD confinement potential.  The features of this $g$-factor correction can be explained by a second order perturbation calculation.  In regions far from the ``hot spots'' of SOC induced anti-crossings, the correction $\delta g$ is rather small.  However, near the critical fields and zero detuning, $\delta g$ could be notable, and the accumulated phase correction could be significant.  One example that the phase error can reduce the transport fidelity is further discussed in Sec.~\ref{subsec:SV-Entanglement}.

\subsection{\label{subsec:SV-Entanglement}Decoherence caused by valley-dependent phase error}

In a Si DQD, valley states could mix with the spin states directly because of spin-orbit coupling \cite{yang2013NatComm}, although this coupling tends to be quite small and is not considered in the present study.  Nevertheless, spin and valley degrees of freedom can indeed mix here because of the presence of a transverse magnetic field gradient and the fact that in a double quantum dot, valley eigenstates are generally different in the two dots.  Furthermore, with electron $g$-factors generally different in different valley states, spin and valley states can mix and entangle even when the two are not directly coupled.  When the information is stored in the spin space, a valley-insensitive spin read-out corresponds to a mathematical operation on the mixed spin-valley density matrix of tracing over the valley states. In this case, a small phase difference between different valley-eigenstates may have a significant impact on spin coherence.  Specifically, if valley and spin states are mixed/entangled, tracing over the valley degree of freedom could cause spin state to become mixed.  Below we present an example where spin coherence is lost because of spin-valley mixing.

Assuming orbital excitation is suppressed within each valley, we can focus on the spin and valley degrees of freedom, so that a pure electron state can be expressed as
\begin{equation}
|\psi_{VS}\rangle=a|+,\uparrow\rangle+b|+,\downarrow\rangle+c|-,\uparrow\rangle+d|-,\downarrow\rangle\,.\label{eq:psiVS}
\end{equation}
The reduced spin density matrix can then be obtained as $\rho_{spin}=\rm{Tr}_V(|\psi_{VS}\rangle\langle\psi_{VS}|)$.  If the two degrees of freedom are separable, the trace here over the valleys would not affect the spin state.  On the other hand, if the two are coupled, and the spin splitting is valley dependent, the trace above would in general be a mixed state, meaning that any initial spin coherence would be at least partially lost.

\begin{figure}
	\includegraphics[width=1\columnwidth]{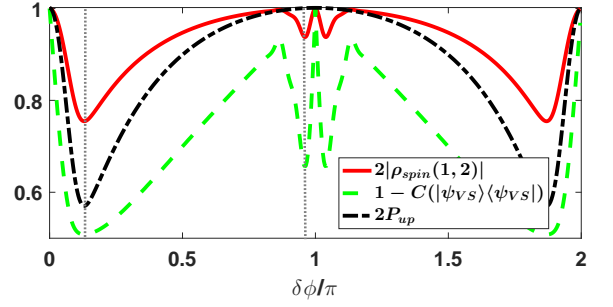}
	
	\caption{\label{fig:P_phi}(color online) Several indicators of spin states as a function of $\delta\phi$.
		The parameters are chosen as $|\Delta_{L}|=60\ {\rm \mu eV}$, $|\Delta_{R}|=30\ {\rm \mu eV}$,
		$t_{C}=45\ {\rm \mu eV}$, $E_{z}=40\ {\rm \mu eV}$, $E_{x}=1.6\ {\rm \mu eV}$,
		operation time $T=10\ {\rm ns}$, detuning is changing from $-0.2\ {\rm meV}$
		to $0.2\ {\rm meV}$. }
\end{figure}

There are several indicators that can be used to evaluate the information loss.  The spin up probability $P_{up}$ is a good measure for classical information (population) loss, while the off-diagonal element of $\rho_{spin}$ is a good measure of the spin coherence or superposition. The SV entanglement $C$ (measured by ``concurrence \citep{wootters1998PRL}'' of $|\psi_{VS}\rangle$) is another indicator, because if $|\psi_{VS}\rangle$ is an entangled state (for the spin and valley degrees of freedom of a single electron), $\rho_{spin}$ would become mixed.  In Fig.~\ref{fig:P_phi}, we plot these indicators as a function of the inter-dot valley phase difference $\delta\phi$, with an initial superposition of the two lowest-energy states $|\psi_{ini}\rangle \approx |L\rangle\otimes| - \rangle \otimes \frac{1}{\sqrt{2}} \left(|\uparrow\rangle + |\downarrow\rangle\right)$. With necessary modifications (re-scale), we normalize all the indicators so that they are equal to $1$ for the initial state, and drop below $1$ if there is coherence or probability loss. Clearly, $2|\rho_{spin}(1,2)|$ and $1-C(|\psi_{VS}\rangle\langle\psi_{VS}|)$ basically follow the same pattern in Fig.~\ref{fig:P_phi}, confirming the positive correlation between SV entanglement and spin coherence loss.

\begin{figure}
	\includegraphics[width=1\columnwidth]{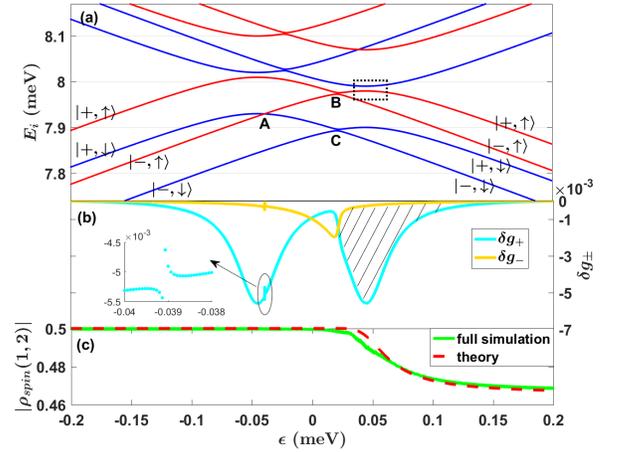}
	
	\caption{\label{fig:SVphase}(color online) (a) Energy diagram of a single-electron Si DQD when $\delta\phi=0.964\pi$, corresponding to one of the spin fidelity dip shown Fig.~\ref{fig:P_phi}. (b) Corrections to $g$-factors for different valley states. (c) Spin coherence after tracing over the valley degree of freedom.  Here the dashed red curve is the result of Eq.~(\ref{eq:rho12}), while the solid green curve is obtained via a full numerical simulation. }
\end{figure}

Figure \ref{fig:P_phi} shows two minima in spin coherence loss or spin transfer fidelity, near $\delta \phi = 0$ and $\delta \phi = \pi$. The second, at $\delta\phi/\pi=0.964$, is the more interesting.  With the two dots having a $\pi$ phase shift in valley phases, the ground valley and first excited eigenstates in one dot becomes almost flipped in the other.  Under such a condition, the anti-crossing between the ground valley states will be strongly suppressed. Consequently, during a detuning sweep an electron in the ground state of one dot will have a finite probability to end up in the excited valley state in the other dot through the Landau-Zener process.  The off-diagonal element of $\rho_{spin}$ (coherence) after such a interdot transport can be approximately expressed as (see Appendix~\ref{sec:SV mixing})
\begin{equation}
|\rho_{spin}(1,2)|=\frac{p}{2}+\frac{1-p}{2} \exp\left\{i\Phi \right\}, \label{eq:rho12}
\end{equation}
where $p$ is the diabatic transition probability at anti-crossings ``B'' and ``C'', and the accumulated phase $\Phi$ is dependent on the effective $g$-factors for the ``$+$'' and ``$-$'' valley states as
\begin{equation}
\Phi=\frac{1}{\hbar}\mu_{B}B_{z}\int_{t_{0}}^{t}[g_{+}(\tau)-g_{-}(\tau)]d\tau \,.
\label{eq:Phi}
\end{equation}
In Fig.~\ref{fig:SVphase}(b), we plot the DQD corrections to the $g$-factors in the different valleys, with $\delta g_\pm=\frac{g_\pm-g_{\pm,s}} {g_{\pm,s}}$, which are benchmarked against the single-dot values $g_{\pm,s}$. The most significant correction to $g_{+}$ comes from the SO mixing shown in the dotted box in Fig.~\ref{fig:SVphase}(a), with a similar correction coming on the negative detuning side as well.

With the corrections to the $g$-factors different for the two valley eigenstates, a phase error would accumulate if the electron has finite probability to be in each of the valley states, as shown in Eq.~(\ref{eq:rho12}) and (\ref{eq:Phi}). Such a phase difference, highlighted in the shadowed area in Fig.~\ref{fig:SVphase}(b), is then reflected in a loss of coherence in the final spin state after the transfer protocol.  In Fig.~\ref{fig:SVphase}(c) we show spin coherence $\rho_{spin}(1,2)$ as a function of detuning, which in turn is a function of time in our spin transfer protocol.  The numerical result shows that a small correction on $g_{\pm}$ ($<0.5\%$) can cause a significant coherence loss ($\sim10\%$).  To validate our theoretical formula~(\ref{eq:rho12}), we also numerically simulate the dynamics of $\rho_{spin}(1,2)$, as represented by the solid green line in Fig.~\ref{fig:SVphase}~(c).  The consistency between the two curves shows that our theory described in Appendix~\ref{sec:SV mixing} successfully captures the main feature of the spin dynamics.

The spin coherence loss in this example started at the valley anti-crossings at ``B'' and ``C'', where the electron state evolves into a SV entangled state through Landau-Zener transitions. The subsequent evolution with different $g$-factors for different valley states means that the electron spin would acquire a valley-dependent phase.  Consequently, a valley-insensitive readout scheme, which effectively trace out the valley degree of freedom, would lead to loss of coherence in the spin state.  A valley-projective measurement would eliminate this error by collapsing the state in Eq.~(\ref{eq:psiVS}) onto a pure spin state in a particular valley.  As long as the $g$-factor in that valley is known, the spin phase can be recovered accurately.

The calculation here shows that even a pure valley anticrossing, such as ``B'' and ``C'', could affect the electron spin state.  Due to the extra valley degree of freedom and the relatively small valley splittings, there are usually multiple relevant anti-crossings in a Si DQD. In Appendix~\ref{sec:MoreAC}, we make a more thorough exploration of various types of anti-crossings, including those that may cause errors in the spin transport, and those that may occur in a wide range of realistic parameters.

In summary, we have studied spin coherence loss due to spin-valley mixing/entanglement and valley-dependent $g$-factor during spin transport through a Si DQD. Our results show that even when spin population is preserved in a process, spin coherence (superposition) could be lost due to, for example, spin-valley mixing and valley-dependent spin splitting.  In other words, while the classical information (population or probability) could be faithfully transported, the quantum information (coherence or superposition) could still be lost along the way.

\section{Spin transport in a double dot: Spin flip errors \label{sec:SpinFlipError}}

With multiple anti-crossings present in the electron spectrum of a Si DQD as interdot detuning is swept, unwanted spin flip, whether through spin relaxation or Landau Zener processes, could lead to significant errors during spin transport.  In this section, We examine spin flip channels due to spin-orbit coupling and inhomogeneous magnetic field, and propose a scheme to reduce spin flip by utilizing an LZ transition to guide the initial state to a state that suffers less decoherence, and restore it afterward through another LZ transition.

\subsection{\label{subsec:SOCflip}spin orbit coupling induced spin flip error}

\begin{figure}
\includegraphics[width=1\columnwidth]{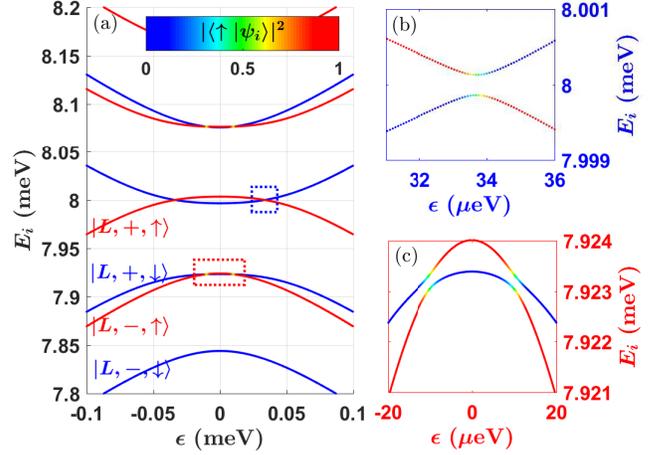}

\caption{\label{fig:SOI}(color online) Energy diagram of DQD and anti-crossings
caused by SV mixing. The parameters are chosen as $\delta\phi=\pi/3$,
$|\Delta_{L}|=|\Delta_{R}|=50\ {\rm \mu eV}$, $t_{C}=72.5\ {\rm \mu eV}$,
$E_{z}=40\ {\rm \mu eV}$, $E_{x}=0\ {\rm \mu eV}$, $S_{1}=S_{2}=0.2\ {\rm \mu eV}$,
operation time $T=8\ {\rm ns}$, detuning is changing from $-0.1\ {\rm meV}$
to $0.1\ {\rm meV}$. The color of the curves represent the spin status
of the eigen-states. The minimum gap is about $0.268\ {\rm \mu eV}$
in (b), and $0.283\ {\rm \mu eV}$ in (c), which are close to the
data measured in Ref. \citep{hao2014NatComm}. The spin fidelity of
the transport (not shown in the figure) is 86.6\% for an initial state
$|L,-,\downarrow\rangle$ and 98.5\% for an initial state $|L,+,\uparrow\rangle$.}
\end{figure}

In this subsection, we investigate the impact of SOC on spin flip without considering the effect of inhomogeneous magnetic field ($E_{x}=0$). In a sweep of interdot detuning, transitions to unwanted states mainly arise from the LZ transitions at the anti-crossings, some of which enables spin flip.
Typically, SOC in silicon is weak, so that almost all SOC-induced anti-crossings are approximately reduced to crossings. However, because of the presence of the valleys in a Si DQD, there is a special regime where the LZ transition can cause significant spin flip error in the transport.

Figure \ref{fig:SOI}~(a) shows a typical energy diagram of a Si DQD.  Two types of anti-crossings are marked with red and blue rectangular box. In the blue box, which is enlarged in subplot (b), the anti-crossing is away from zero detuning and between states that have different locations (left and right here). The energy difference here depends strongly on the interdot detuning near this anti-crossing.
Typically, at such an anti-crossing $a^{2}/\hbar$ is much smaller than $d|E_{m}-E_{n}|/dt$, so that $P_{D}\approx 1$ and the anti-crossing becomes roughly a crossing.  In contrast, in the red box, which is enlarged in subplot (c), the anti-crossing occurs near zero detuning, and is between different valley states in the same dot. Here the two energy levels change slowly relative to each other because they are dominated by different valley states in the same quantum dot, which allows the possibility of $a^{2}/\hbar$ being comparable to $d|E_{m}-E_{n}|/dt$. With this anti-crossing existing in a much broader range of detuning, it could cause more significant spin flip during spin transport.

A series of numerical simulations with the same set of parameters except the initial states confirms the qualitative analysis above.  For an initial state prepared in the fourth lowest energy level (approximately $|L,+,\uparrow\rangle$ initially), it passes through two anti-crossings away from zero detuning as $\epsilon$ is swept from negative to positive. Numerical results show that the probability of keeping spin up is about $98.5\%$. In comparison, for an initial state in the second lowest energy level (approximately $|L,-,\uparrow\rangle$ initially), it passes through two anti-crossings near zero detuning as $\epsilon$ is swept. The probability of keeping spin up is sharply decreased to $86.6\%$.

In short, most anti-crossings in a Si DQD cannot cause significant spin flip during spin transport because of the weak SOC in silicon. The only case that deserves special attention is the anti-crossings near zero detuning, where pairs of valley states have energy differences that only depend on interdot detuning weakly.  The wide range of nearly parallel states, while possibly allowing weaker charge-noise induced dephasing~\citep{shi2012PRL,koh2012PRL,russ2018PRL},
results in much larger probabilities of spin flip.

\subsection{\label{subsec:NMflip}Inhomogeneous magnetic field induced spin flip}

\begin{figure}
\includegraphics[width=1\columnwidth]{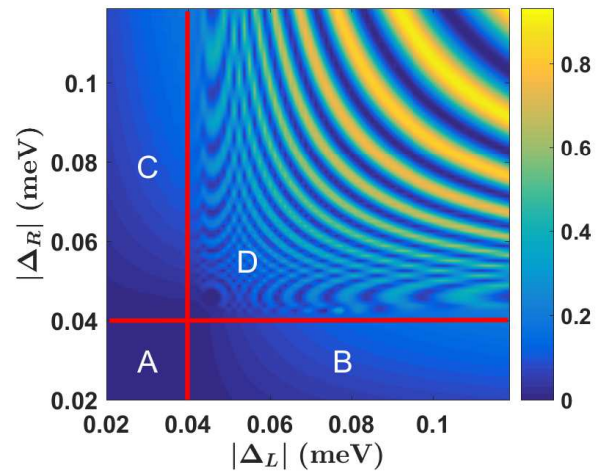}

\caption{\label{fig:FvsDLDR}(color online) An example of spin transport infidelity as a function of the valley splittings in the left and right dots in the presence of a micromagnet. The system parameters are chosen as $\phi_{L}=-\pi/3$, $\phi_{R}=\pi/3$, $t_{C}=70\ {\rm \mu eV}$, $E_{z}=40\ {\rm \mu eV}$, $E_{x}=2\ {\rm \mu eV}$, and detuning is changed from $-0.2\ {\rm meV}$ to $0.2\ {\rm meV}$ in $T=10$ ns.}
\end{figure}

While SOC usually has a limited impact on spin flip in a Si DQD, the inhomogeneous magnetic field from a micromagnet can connect different
spin states strongly and allow electric dipole spin resonance~\citep{tokura2006PRL,mi2018Nature,samkharadze2018Science,hu2012PRB,pioro2008NatPhys}.
Here we focus on an inhomogeneous field with a gradient in the $x$-direction (interdot axial direction). Comparing to the strength of SOC ($S_{1}\sim0.2\ {\rm \mu eV}$ \citep{hao2014NatComm}), the coupling between spin states caused by $\pm B_{x}$ could be tuned to larger than $1\ \mu$eV \citep{benito2017PRB}. As a result, most anti-crossings caused by a transverse field gradient can induce relatively fast spin flips. Here, the main issues are the conditions for the formation of such anti-crossings, and the types of anti-crossings that lead to fastest spin flips.

Figure \ref{fig:FvsDLDR} shows spin infidelity $1-F_{spin}=|P_{\sigma,fin}-P_{\sigma,ini}|$ after the transport in the parameter space of single dot valley splittings, where $P_{\sigma,ini}$ and $P_{\sigma,fin}$ indicate the spin population for initial state and final state respectively.  We choose a particular Zeeman splitting $E_{z}=40\mbox{\ensuremath{\mu}eV}$, then change the valley splittings in left and right dots. The initial state is always chosen as the spin excited but orbital ground state at $\epsilon\ll-t_{C}$.  The relations of three important energy scales $|\Delta_{L}|$, $|\Delta_{R}|$, and $E_{z}$ divide the given parameter space into four regions as we marked in Fig.~\ref{fig:FvsDLDR} (The detailed energy diagrams for these four cases are plotted in Appendix \ref{sec:Energy Diagram}). The spin dynamics is dramatically different in these regions.

In the high-field region ``A'', both $|\Delta_{L}|$ and $|\Delta_{R}|$ are smaller than the Zeeman energy $E_{z}$. With state $|-,\uparrow\rangle$ always having higher energy than $|+,\downarrow\rangle$, there is no crossings or anti-crossings in the energy diagram when interdot detuning $\epsilon$ is swept from negative to positive. Consequently, there is no significant spin flip error in region ``A'' when an electron spin is transported.

In the intermediate-field region ``B'' and ``C'', one of the valley splittings is smaller than $E_{z}$, while the other is larger. As a result,
an anti-crossing appears in the energy diagram when detuning is varied. For example, in region ``B'', where $|\Delta_{L}|>E_{z}>|\Delta_{R}|$, the first excited state in left dot is a spin excited state, while in the right dot it is valley excited state.
When interdot detuning is varied from negative to positive and the electron moves from left to right, the first excited state has to change, indicating the presence of a level anti-crossing (due to $E_{x}$ and/or SOC).  Such an anti-crossing is reflected in the numerical results in Fig.~\ref{fig:FvsDLDR}, which shows that there is a small but notable spin flip error in regions ``B'' and ``C'' due to the anti-crossings between $|-,\uparrow\rangle$ and $|+,\downarrow\rangle$.

Lastly, in the low-field region ``D'', where $|\Delta_{L}|>E_{z}$ and $|\Delta_{R}|>E_{z}$, it is possible to form two anti-crossings in the energy diagram between $|-,\uparrow\rangle$ and $|+,\downarrow\rangle$ (See the energy diagram Appendix \ref{sec:Energy Diagram}). The interference between these two anti-crossings can either enhance or reduce spin flip, as is shown in Fig.~\ref{fig:FvsDLDR} in the interference pattern. Given different $|\Delta_{L}|$ and $|\Delta_{R}|$, the two anti-crossings would form at different detunings, so that the dynamical phase accumulated between the two anti-crossings would be different when the detuning is swept, leading to $\Delta_{L/R}$-dependent interference pattern in the figure. Spin flip error in this region can be large because the unwanted transitions can be amplified by the interference.

In summary, we have explored the conditions for the formation of anti-crossings, and identified four parameter regions with respect to the valley splittings $|\Delta_{L}|$ and $|\Delta_{R}|$, and the Zeeman splitting $E_{z}$. In the high field region, when $|\Delta_{L}|<E_{z}$ and $|\Delta_{R}| < E_{z}$, no anti-crossing forms, and spin flip probability is minimized. In the region of intermediate field, when $|\Delta_{L}| > E_{z} > |\Delta_{R}|$ or $|\Delta_{R}| > E_{z} > |\Delta_{L}|$, one anti-crossing appears, which may cause spin flip. In the low field region, when $|\Delta_{L}|>E_{z}$ and $|\Delta_{R}| > E_{z}$, two anti-crossings form, and spin flip could be significantly enhanced during spin transport by interference between the two anti-crossings.

\subsection{\label{subsec:Suppress_Spin_Flip}Suppression of spin flip errors
	by LZ transitions}

\begin{figure}
	\includegraphics[width=1\linewidth]{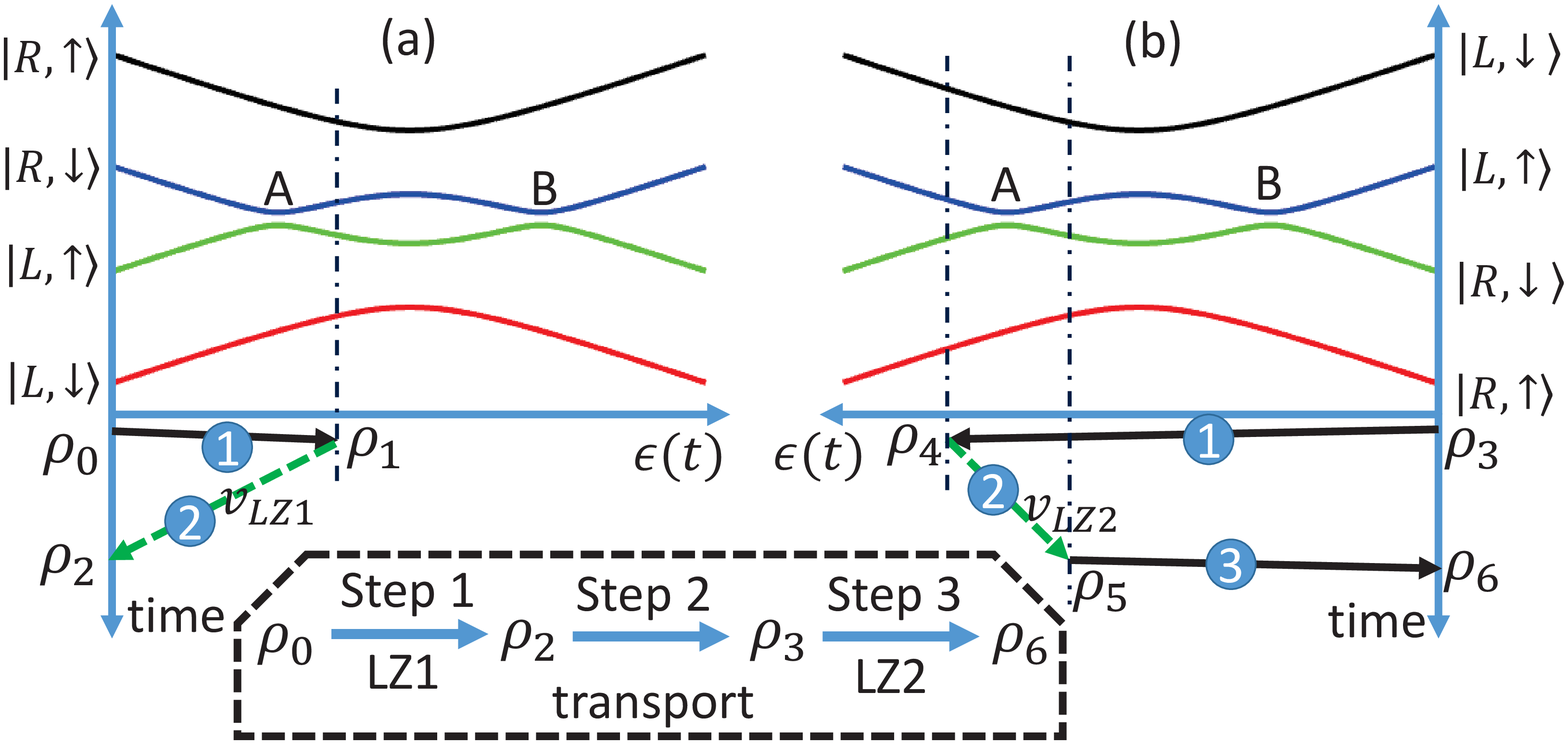}
	
	\caption{\label{fig:LZoperation}(color online) Scheme of using LZ operations
		to suppress spin relaxation}
\end{figure}

As we discussed above, spin flip, whether caused by unwanted transitions or spin relaxation\citep{raith2011PRB,Zhao2018SR,benito2017PRB}, could be an important error during spin transport. In this subsection, we propose a scheme to reduce spin flip errors probabilistically by utilizing LZ transitions. The basic idea can be traced back to probabilistic quantum error correction by weak measurement \citep{koashi1999PRL}. However, in our scheme, all operations are performed within the DQD system, in contrast to the example given in Ref.~\onlinecite{koashi1999PRL}, where extra auxiliary qubits are needed.

For simplicity, we consider an example involving only spin states and two orbital states $|L\rangle$ and $|R\rangle$, so that Hamiltonian~(\ref{eq:HSVO}) is reduced to $H = \epsilon\tau_{z} + t_{C}\tau_{x} + E_{z}\sigma_{z} + E_{x}\tau_{z}\sigma_{x}$.  The operators $\tau_{x,y,z}$ and $\sigma_{x,y,z}$ are Pauli matrices in the orbital and spin spaces, respectively. Such a configuration is used in recent experiments \citep{benito2017PRB,mi2018Nature} to realize spin-photon coupling.  A typical energy diagram is plotted in Fig.~\ref{fig:LZoperation} with two anti-crossings caused by an inhomogeneous magnetic field.

Consider now the transfer of the following unknown state in the left dot,
\begin{equation}
|\psi_{0}\rangle=|L\rangle\otimes(\alpha|\uparrow\rangle+\beta|\downarrow\rangle)\quad(|\alpha|^{2}+|\beta|^{2}=1),
\end{equation}
to the right dot.  The ground spin state $|\downarrow\rangle$ has the lowest energy, suffers no spin relaxation, and generally experiences less spin flip error.  We can in principle take advantage of these favorable properties by ``pushing'' the initial state close to the ground state before the transport and then try to recover the unknown state by another LZ operation after the transport.  Qur protocol that uses the following three steps.

\textbf{Step 1:} Probabilistically prepare $|\psi_{0}\rangle$ to a state which may suffer less spin flip.  This can be realized by rapidly sweeping detuning $\epsilon$ over anti-crossing ``A'' and then slowly shifting it back with a controlled LZ velocity $v_{{\rm LZ1}}$, as shown in Fig.~\ref{fig:LZoperation}~(a). The ground state $|L,\downarrow\rangle$ is not affected by such an operation, while the spin excited state $|L,\uparrow\rangle$ would be split, so that the initial state is modified: $|\psi_{0}\rangle\rightarrow\sqrt{p}\alpha|L,\uparrow\rangle+\sqrt{1-p}\alpha|R,\downarrow\rangle+\beta|L,\downarrow\rangle$, where $p=\exp\left(-\frac{2\pi E_{x}^{2}/\hbar}{v_{{\rm LZ1}}}\right)$ is the diabatic transition probability when shifting back. One then perform a charge measurement to distinguish orbital states $|L\rangle$ and $|R\rangle$.  If orbital state $|L\rangle$ is detected, the overall electron state would collapse to
\begin{equation}
|L\rangle\langle L|\otimes\rho_{2}=|L\rangle\langle L|\otimes\frac{1}{N}\left[\begin{array}{cc}
p|\alpha|^{2} & \sqrt{p}\alpha\beta^{*}\\
\sqrt{p}\alpha^{*}\beta & |\beta|^{2}
\end{array}\right] \,, \label{eq:LZ1}
\end{equation}
where $N$ is the normalization factor.  The density matrix here still contains the information from the original state, convoluted with the LZ transition probability $p$.

\textbf{Step 2:} Perform the intended electron transport with a possible spin flip error, so that $|L\rangle\langle L| \otimes \rho_{2} \rightarrow |R\rangle \langle R|\otimes\rho_{3}$.  Here we have included any spin flip as an amplitude damping process (see Appendix~\ref{sec:Supp_Suppress}), so that $\rho_3$ takes the form
\begin{equation}
\rho_{3}=\frac{1}{N}\left[\begin{array}{cc}
\Gamma_{A}^{2}p|\alpha|^{2} & \Gamma_{A}\sqrt{p}\alpha\beta^{*}\\
\Gamma_{A}\sqrt{p}\alpha^{*}\beta & (1-\Gamma_{A}^{2})p|\alpha|^{2}+|\beta|^{2}
\end{array}\right].\label{eq:Relaxation}
\end{equation}
Using spin relaxation as an example, $\Gamma_{A}=\exp[-\int_{0}^{t}\gamma_{A}(\tau)d\tau]$ with $\gamma_{A}(\tau)$ the spin relaxation rate \citep{raith2011PRB,Zhao2018SR,benito2017PRB}.  Such a spin flip error can also be caused by an unwanted transition in the transport.  A detailed study for that case is given in Appendix~\ref{sec:Supp_Suppress}.

\textbf{Step 3:}
Recover the initial spin state by performing a similar LZ operation as shown in Fig.~\ref{fig:LZoperation}~(b) and a charge detection.  The operation ``LZ2'' can reduce the probability of state $|\downarrow\rangle$ (a detailed discussion is in Appendix \ref{sec:LZ2}), thus helping to recover the initial state. In analogy to Eq.~(\ref{eq:LZ1}), the state in Eq.~(\ref{eq:Relaxation}) becomes $|R\rangle\langle R|\otimes\rho_{6}$ if the charge measurement decides that the electron has been transported to the right dot.  The spin density matrix after this charge projection is
\begin{equation}
\rho_{6}=\frac{1}{N^{\prime}}\left[\begin{array}{cc}
\Gamma_{A}^{2}p|\alpha|^{2} & \Gamma_{A}\sqrt{pp^{\prime}}\alpha\beta^{*}\\
\Gamma_{A}\sqrt{pp^{\prime}}\alpha^{*}\beta & (1-\Gamma_{A}^{2})pp^{\prime}|\alpha|^{2}+p^{\prime}|\beta|^{2}
\end{array}\right],\label{eq:FinalState}
\end{equation}
with a diabatic LZ probability $p^{\prime}=\exp\left(-\frac{2\pi E_{x}^{2}/\hbar}{v_{{\rm LZ2}}}\right)$ determined by $v_{{\rm LZ2}}$, and $N^{\prime}$ is a normalization factor. Comparing to the initial state $|\psi_{0}\rangle\langle\psi_{0}|=|L\rangle\langle L|\otimes\rho_{0}$, the difference between spin states $\rho_{0}$ and $\rho_{6}$ can be measured by the trace distance $D(\rho_{0}, \rho_{6})=\frac{1}{2} \sum |\lambda_{i}|$, where $\lambda_{i}$ are the eigenvalues of the matrix $\rho_{0}-\rho_{6}$. Since $v_{{\rm LZ1}}$ and $v_{{\rm LZ2}}$ are controllable, a perfect restoration is possible \citep{koashi1999PRL}. For example, if we take $p^{\prime}=\Gamma_{A}^{2}p$,
\begin{equation}
D(\rho_{0},\rho_{6})\propto\frac{1}{2}(1-\Gamma_{A}^{2})p|\alpha|^{2}.\label{eq:Drho0rho3}
\end{equation}
The trace distance $D\rightarrow0$ as $p\rightarrow0$, so that the final spin state $\rho_{6}$ can approach the initial unknown state $\rho_{0}$ infinitely closely.

\begin{figure}
	\includegraphics[width=1\linewidth]{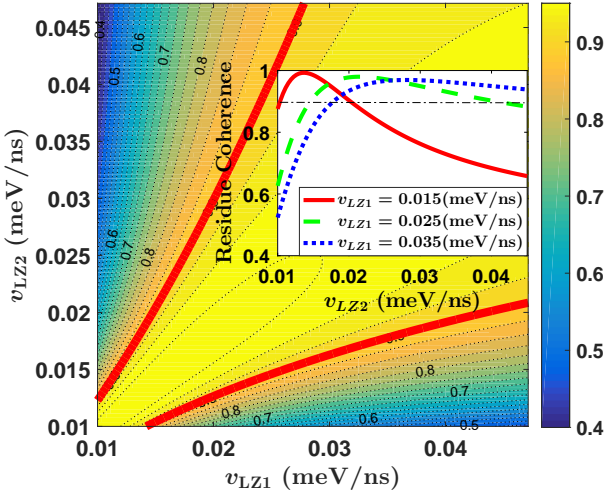}
	
	\caption{\label{fig:RecP}(color online) Residue coherence $\frac{|\rho_{6}(1,2)|}{|\rho_{0}(1,2)|}$
		after the transport. The parameters are chosen as $E_{x}=2{\rm \mu eV}$,
		$\Gamma_{A}=0.9$. Initial state is $\frac{1}{\sqrt{2}}|L\rangle\otimes(|\uparrow\rangle+|\downarrow\rangle)$.}
\end{figure}

To achieve a perfect recovery, $v_{LZ1}$ and $v_{LZ2}$ must be precisely controlled, and a good estimate of $\gamma_A$ is required. However, the numerical study in Fig.~\ref{fig:RecP} reveals that our protection scheme can work even in non-ideal cases.  The numerical results show that the protection is not very sensitive to the precision of $v_{{\rm LZ1}}$ and $v_{{\rm LZ2}}$. Indeed, spin coherence is better preserved with this protection procedure than without when $(v_{{\rm LZ1}},v_{{\rm LZ2}})$ is in a wide range highlighted between the two red-solid lines in Fig.~\ref{fig:RecP} (where $\frac{|\rho_{6}(1,2)|}{|\rho_{0}(1,2)|}=0.9$) . The inset of the figure clearly shows that for a given $v_{{\rm LZ1}}$, there is a wide range to choose $v_{{\rm LZ2}}$ to achieve a positive protection effect, and the final state can be recovered almost identical ($\frac{|\rho_{6}(1,2)|}{|\rho_{0}(1,2)|}\approx 1$) to the initial state when $v_{{\rm LZ1}}$ and $v_{{\rm LZ2}}$ are chosen properly.

While the objective of this scheme is to protect against spin flip errors, it is important to keep track of the phase of the spin qubit.  A simple method to track the phase shift in such an experiment is to perform a test run by transporting a known initial state and determine the phase shift for a set of given parameter by measuring the final state. Furthermore, with knowledge of the DQD parameters ($t_C$, $E_z$, $\epsilon$ etc.), the phase shift can also be numerically computed.

In summary, we propose a scheme to protect the transported spin qubit from spin flip errors by utilizing LZ transitions. With the multiple level anti-crossings in a Si DQD, our proposal here provides a good example of utilizing these anti-crossings to achieve high-fidelity transport.

\section{\label{sec:Conclusion}Conclusion}

In this paper, we have studied the spectrum and dynamics of a single
electron in a silicon DQD. We first clarify the spectrum of the electron
charge motion, and investigate the charge dynamics (valley and orbital)
as the electron is driven between the two dots. We identify the phase
difference of the valley-orbit coupling matrix elements in the two
dots as a key parameter in determining the tunnel coupling between
the two dots, and propose several schemes to detect this phase difference
via transport through the DQD, or via charge sensing after pulsing
the DQD through the anti-crossings near zero detuning. The LZ transitions
or even LZS interference during the detuning sweep lead to different
charge distribution between the two dots as we change the pulse sequence,
which allow us to calculate the valley phase difference. We derive
expressions for the valley phase difference under different conditions
and discuss feasibility of these schemes within the current experimental
technologies.

Besides the valley orbital dynamics, we focus on the spin dynamics
and study several factors that may cause phase errors and spin flip errors. (1) We find
the effective $g$-factor will be modified in a DQD, which causes
a notable accumulated dynamical phase error. (2) We investigate the relationship
between spin transfer fidelity and the valley phase difference, and
analyze loss of spin purity caused by SV mixing. We show an example
where the purity of the spin state is lost even though the spin population
is faithfully transported. (3) In most cases, SOC
caused spin flip remains minimal due to the small SOC in silicon.
However, we identify a special type of anti-crossing near zero detuning,
which contains a wide region (in detuning) where state mixing is significant,
so that considerable spin flip can occur as a result. (4) Recognizing
the importance of the various types of anti-crossings to spin and
charge transfer fidelity, we identify four different regions in the
parameter space defined by valley splitting and Zeeman splitting,
in which spin transfer fidelity has distinct dependences on the parameters
and the resulting anti-crossings. (5) We propose a scheme to probabilistically
suppress the spin relaxation by using LZ transitions.

To summary, we try to make a thorough examination of quantum coherent
electron transport in a silicon DQD, and study many possible factors
that may affect the transport. We believe all these factors are closely
related to current or future experiments. It is our hope that our
study could be helpful to various transport experiments.
\begin{acknowledgments}
We acknowledge financial support by US ARO (W911NF1710257).
\end{acknowledgments}

\appendix

\section{\label{sec:Calculation-geff}Calculation of effective $g$-factor}

The electron spin splitting, represented by the Landé $g$-factor,
is affected by the orbital spectrum through the SOC. Here we examine
how the double dot confinement potential modifies the $g$-factor,
which in turn modifies the phase of the excited spin state as an electron
is transported through the double dot.

The total Hamiltonian for an electron confined in a DQD is $H=H_{DQD}+H_{Z}+H_{SO}+H_{V}$.
The spin part is governed by $H_{Z}=\frac{1}{2}g\mu_{B}B\sigma_{z}$,
with $g$ here the bulk $g$-factor in Si. $H_{V}$ represents valley-orbit
coupling, and is a 2 by 2 matrix with off-diagonal elements $\Delta_{D}=|\Delta_{D}|e^{i\phi_{D}}$
($D=L,R$). The SO interaction is given by
\begin{equation}
H_{SO}=\frac{\alpha_{BR}}{\hbar}(\sigma_{x}\pi_{y}-\sigma_{y}\pi_{x})+\frac{\alpha_{D}}{\hbar}(\sigma_{y}\pi_{y}-\sigma_{x}\pi_{x}),
\end{equation}
where $\alpha_{D}$ and $\alpha_{BR}$ are the strengths of Dresselhaus
and Bychkov-Rashba SOC, respectively \citep{dresselhaus1955PR,bychkov1984JPC}.
The matrix elements of $H$ can be computed from a modeled Hamiltonian
$H_{DQD}=T+V+\epsilon\frac{x}{d}$, where $T=\frac{\mathbf{\pi}^{2}}{2m^{*}}$
is the kinetic energy, $V(x,y)=\frac{1}{2}m^{*}\omega_{0}^{2}[(|x|-d)^{2}+y^{2}]$
is the potential energy, and $\epsilon\frac{x}{d}$ is the detuning
between two dots from an external electric field along the inter-dot
axis. Here $m^{*}$ is the effective mass of the electron, $\mathbf{\pi}=\mathbf{p}+e\mathbf{A}$
is the kinetic momentum operator, and $\mathbf{A}=B(-y/2,x/2,0)$
is the vector potential of the applied magnetic field.

For a single quantum dot, the eigen-states in the absence of $H_{SO}$
are the Fock-Darwin states
\begin{equation}
\psi_{nl\sigma v}=C\rho^{|l|}e^{-\rho^{2}/2}L_{n}^{|l|}(\rho^{2})e^{il\phi}\varphi_{\sigma}\varphi_{v},\label{eq:FD}
\end{equation}
where $n,l,\sigma$ are the primary quantum number, the orbital angular
momentum quantum number, and the spin quantum number, while $\varphi_{v}$
represents the valley states $z$ or $\bar{z}$. $\rho=\sqrt{x^{2}+y^{2}}/l_{B}$
is the in-plane radius for the electron, with $l_{B}^{2}=l_{0}^{2}\sqrt{1+B^{2}e^{2}l_{0}^{4}/4\hbar^{2}}$
and $l_{0}=\sqrt{\hbar/m\omega_{0}}$ the effective confinement length
including the effective confinement produced by $B$ field. $\phi$
is the angle in polar coordinate defined as $\tan\phi=y/x$. $L_{n}^{|l|}$
is the associated Laguerre polynomials.
In a double quantum dot, the orbital eigenstates can be expanded on
a basis of shifted single-dot Fock-Darwin states $\psi_{nl\sigma v}^{L}=\psi_{nl\sigma v}(x+d,y)\exp(\frac{iyd}{2b^{2}})$
and $\psi_{nl\sigma v}^{R}=\psi_{nl\sigma v}(x-d,y)\exp(\frac{-iyd}{2b^{2}})$.
A set of orthogonal basis can be constructed from the single-dot states
\citep{culcer2010PRB2}. In Eq.~(\ref{eq:HSVO}) for valley-orbit
dynamics, only the ground orbital $s$ states are included. However,
the $H_{SO}$ interaction may also couple the ground orbital $s$
states with higher orbital $p$ states. To make it more accurate here,
we will also include the orbital $p$ states $|\psi_{0\pm1\sigma}^{L,R}\rangle$.
So, the total Hamiltonian can be expressed in a $24\times24$ matrix
in the basis $\{\psi_{nl\sigma v}^{L},\psi_{nl\sigma v}^{R}\}$ ($n=0$;
$l=0,\pm1$; $\sigma=\uparrow,\downarrow$; $v=z,\bar{z}$). Diagonalizing
the total Hamiltonian, one can obtain the lowest two eigen-energies
$\epsilon_{g,\downarrow}$ and $\epsilon_{g,\uparrow}$ corresponding
to ground orbital state with two different spin states.

\section{\label{sec:SV mixing} Analysis of spin coherence loss due to SV mixing}

In this Appendix, we provide a detailed analysis of the spin-valley mixing during spin transport, and derive Eqs.~(\ref{eq:rho12}) and (\ref{eq:Phi}).

For a general spin-valley mixed state given in Eq.~(\ref{eq:psiVS}), the off-diagonal spin density matrix element is $\rho_{spin}(1,2)=ab^{*}+cd^{*}$. In our spin transfer protocol, the initial state is $\psi_{ini} = |L\rangle \otimes |-\rangle \otimes \frac{1}{\sqrt{2}} \left( |\uparrow\rangle + |\downarrow \rangle \right)$, which means initially $a = b = 0$.  Before the anti-crossings ``B'' and ``C'' in Fig.~\ref{fig:SVphase}(a), which are pure valley anti-crossings and are assumed to be swept through at time $t_{0}$, the coefficients $a(t<t_{0})$ and $b(t<t_{0})$ remain zero (i.e. no valley excitation) because the anti-crossing ``A'' does not cause any notable transition.  The gap at ``A'', which is due to direct spin-valley coupling, is extremely small (about 0.35 $\mu$eV) making it more a ``crossing'' rather than an ``anti-crossing'' under any reasonable sweeping speed. This is verified in Fig.~\ref{fig:P_phi}, where the numerically calculated population $P_{up}$ does not decrease at $\delta\phi=0.964\pi$.

Before $t_{0}$, spin phase evolution does not affect the magnitude of the coherence because $|\rho_{spin}(1,2)|=|c(t)d^*(t)|=|c(0)d^*(0)|$.  The picture changes around $t_0$.  The electron undergoes Landau-Zener transitions at valley anti-crossings ``B'' and ``C'' following Eq.~(\ref{eq:LZF}).  Immediately after the anti-crossings ``B'' and ``C'', the electron state evolves into
\begin{eqnarray}
|\psi_{VS}(t_{0})\rangle & = & a(t_{0})|+,\uparrow\rangle+b(t_{0})|+,\downarrow\rangle\nonumber \\
&  & +c(t_{0})|-,\uparrow\rangle+d(t_{0})|-,\downarrow\rangle,
\end{eqnarray}
where $a(t_{0})=b(t_{0})=\sqrt{p/2}$ and $c(t_{0})=d(t_{0})=\sqrt{\frac{1-p}{2}}$ (We have assumed here that $c(t<t_{0}) = d(t<t_{0}) = \frac{1}{\sqrt{2}}$, and have neglected the phase of the electron spin state before this point.  This phase can be easily incorporated in the discussion here if the temporal profile of the detuning sweep is known).

In the evolution for $t>t_0$, the valley-dependent accumulated phase for the electron spin will play an important role. After a period of adiabatic evolution, the state will evolve to $|\psi_{VS}(t)\rangle=\exp\left[-\frac{i}{\hbar}\int_{t_{0}}^{t}H(\tau)d\tau\right]|\psi_{VS}(t_{0})\rangle$,
i.e.,
\begin{eqnarray}
|\psi_{VS}(t)\rangle & = & \sqrt{\frac{p}{2}}\exp\left[-i\int_{t_{0}}^{t}E_{+,\uparrow}(\tau)d\tau\right]|+,\uparrow\rangle\nonumber \\
&+& \sqrt{\frac{p}{2}}\exp\left[-i\int_{t_{0}}^{t}E_{+,\downarrow}(\tau)d\tau\right]|+,\downarrow\rangle\nonumber \\
&+& \sqrt{\frac{1-p}{2}}\exp\left[-i\int_{t_{0}}^{t}E_{-,\uparrow}(\tau)d\tau\right]|-,\uparrow\rangle\nonumber \\
&+& \sqrt{\frac{1-p}{2}}\exp\left[-i\int_{t_{0}}^{t}E_{-,\downarrow}(\tau)d\tau\right]|-,\downarrow\rangle,\label{eq:psivst}
\end{eqnarray}
where $E_{+,\uparrow}$, $E_{+,\downarrow}$, $E_{-,\uparrow}$, and $E_{-,\downarrow}$ are the instantaneous eigen-energies of the corresponding SV states \mbox{$|+,\uparrow\rangle$}, \mbox{$|+,\downarrow \rangle$}, $|-,\uparrow\rangle$, and $|-,\downarrow\rangle$.  After tracing out the valleys (which is equivalent to assuming a valley-insensitive detection method), the off-diagonal spin density matrix element is
\begin{eqnarray}
\rho_{spin}(1,2)&=&a(t)b^{*}(t)+c(t)d^{*}(t) \nonumber\\ & = & \frac{p}{2}\exp\left\{ -\frac{i}{\hbar}\int_{t_{0}}^{t}g_{+}(\tau)\mu_{B}B_{z}d\tau\right\} \nonumber \\
& + & \frac{1-p}{2}\exp\left\{ -\frac{i}{\hbar}\int_{t_{0}}^{t}g_{-}(\tau)\mu_{B}B_{z}d\tau\right\}. \label{eq:ab+cd}
\end{eqnarray}
where $g_{+}(\tau)\mu_{B}B_{z}=E_{+,\uparrow}(\tau)-E_{+,\downarrow}(\tau)$ and $g_{-}(\tau)\mu_{B}B_{z}=E_{-,\uparrow}(\tau)-E_{-,\downarrow}(\tau)$
are the energy gaps between spin up and down states for $+$ and $-$ valleys.

Without considering the DQD correction to the $g$-factors, $g_{\pm}(\tau)\mu_{B}B_{z}$ should both be equal to the bulk Zeeman splitting $E_{z}$, namely $g_{+}=g_{-}$.  As a result, the electron spin would only acquire a global phase factor $\exp\left[-\frac{i}{\hbar}E_{z}(t-t_{0})\right]$, which does not affect the amplitude of the off-diagonal elements $|\rho_{spin}(1,2)|$.

However, the DQD potential does produce a correction on the effective $g$-factor. More importantly, the corrections on $|+\rangle$ and $|-\rangle$ valley states are different, as shown in Fig.~\ref{fig:SVphase}(b).  Consequently, a relative phase between the valleys develops in Eq.~(\ref{eq:ab+cd}), which entangles spin and valley degrees of freedom, and reduces the spin coherence.  Since a global phase does not contribute to the absolute value $|\rho_{spin}(1,2)|$, one can factor out a phase $\exp\left\{ \frac{i}{\hbar}\int_{t_{0}}^{t}g_{+}(\tau)\mu_{B}B_{z}d\tau\right\} $
from Eq.~(\ref{eq:ab+cd}), and obtain formula~(\ref{eq:rho12}) and (\ref{eq:Phi}).  As shown in Fig.~\ref{fig:SVphase}(b), the global phase here is affected by the DQD confinement potential, and the modification would be contained in the phase of $\rho_{spin}(1,2)$, as we have discussed in Sec.~IV.A.

In the derivation above, we have focused on the effect of the valley-dependent accumulated phase. The numerical result in Fig.~\ref{fig:SVphase}~(c) shows that our theoretical prediction Eq.~(\ref{eq:rho12}) is very close to the full numerical simulation, justifying our belief that the valley-dependent phase is the main cause of the spin fidelity loss.  Furthermore, the key to the amplitude reduction in spin coherence is the phase difference between the valleys.  As we discussed in the main text, the average phase accumulation due to the DQD-modified $g$-factor, which is not discussed here but directly affect the phase of spin coherence, is also an important factor in maintaining fidelity of a superposed qubit state.

\section{\label{sec:MoreAC}General presence of SV anti-crossings in a silicon DQD}

\begin{figure}
	\includegraphics[width=1\linewidth]{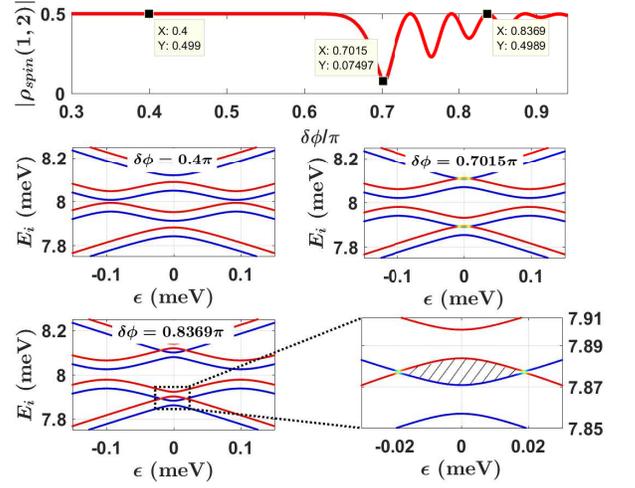}
	
	\caption{\label{fig:Intf}(color online) Residue coherence $|\rho_{spin}(1,2)|$
		after transport as a function of valley phase difference $\delta\phi$.
		For three interesting $\delta\phi$ points, the corresponding energy
		diagrams are plotted. The parameters are $|\Delta_{L,R}|=0.1\;{\rm meV}$,
		$E_{z}=40\;{\rm \mu eV}$, $E_{x}=1.6\;{\rm \mu eV}$, and $t_{C}=45\;{\rm \mu eV}$.}
\end{figure}

In the example we analyzed in Section IV.B, the spin-valley anti-crossing ``A'' is very narrow so that it acts like a crossing, and spin-valley mixing there is caused by valley anti-crossings ``B'' and ``C'' and the valley-dependent $g$-factor.  Here we present another example where spin-valley anti-crossing (from the transverse magnetic field gradient and the dot-dependent valley mixing) has a larger magnitude and cause spin-valley mixing directly.

In this section, we take a closer look at SV anti-crossings in a silicon DQD as a supplement to the case discussed in Sec.~\ref{subsec:SV-Entanglement}, where we show that the SV mixing can cause SV entanglement thus reducing the spin coherence. In Fig.~\ref{fig:SVphase}, the major transitions occur at anti-crossings ``B'' and ``C'', which are purely valley transitions between valley states $|+\rangle$ and $|-\rangle$ of the left and right dots.  The SV energy gap at ``A'' is too small ($\sim 0.35 \mu$eV) to cause notable transitions.

In Sec.~\ref{subsec:SV-Entanglement}, the valley splitting $|\Delta_{L,R}|$ are chosen close to the Zeeman splitting $E_{z}$. In a general
case, $|\Delta_{L,R}|$ may reach $0.1\;{\rm meV}$ in SiGe heterostructures and $0.3-0.8$ meV in MOS structures \cite{yang2013NatComm}. With a larger $|\Delta_{L,R}|$, it seems easier to keep valley splitting larger than the Zeeman splitting, therefore avoiding unwanted SV anti-crossings.  However, as we show here, with valley-orbit phase generally different across a double quantum dot, SV anti-crossings is almost unavoidable, and will happen either at inter-valley or intra-valley transitions.  Furthermore, with coupled dynamics between spin, orbital, and valley degrees of freedom, spin-valley anti-crossings (due to magnetic field gradient here) can be enhanced by interdot mixings.  For example,
as is shown in Fig.~\ref{fig:Intf}, the intra-valley tunneling gap $t_{-}$ is strongly dependent on the valley phase difference $\delta\phi$. When $\delta\phi$ is large, $|t_{-}|$ becomes smaller than $E_z$, so that SV anti-crossings would appear near the intra-valley anti-crossing.  Specifically, with the parameters chosen for Fig.~\ref{fig:Intf},
the intra-valley tunneling gap $|t_{-}|$
is larger than the Zeeman splitting $E_{z}$ when $\delta\phi<0.6\pi$, and there is no SV anti-crossing.
A typical case $\delta\phi=0.4\pi$ is plotted in Fig.~\ref{fig:Intf}.  However, when $\delta\phi$ is closer to $\pi/2$, so that intra-valley tunneling is reduced, anti-crossings around the intra-valley
tunneling gap would develop, as shown in the energy diagram for $\delta\phi=0.7015\pi$
and $\delta\phi=0.8369\pi$.  In addition, the SV anti-crossings are also enhanced here to about 1.5 $\mu$eV, making LZ transitions more likely to happen at each anti-crossing.

The results shown in Fig.~\ref{fig:Intf} indicate that with realistic parameters, SV anti-crossings are quite common in a Si DQD, and could cause coherence loss when a spin qubit is transferred through them.  Since a single atomic step on the interface near (or inside) the QD may cause a significant valley phase difference $\delta\phi$ \citep{tariq2019arXiv} and it is difficult to avoid or control these steps within the current technology, coherence loss caused by transitions at anti-crossings could be an important issue in many cases, and has to be accounted for in a transport experiment.

\section{\label{sec:Energy Diagram}Energy diagrams for different valley splitting
	configuration}

\begin{figure}
	\includegraphics[width=1\columnwidth]{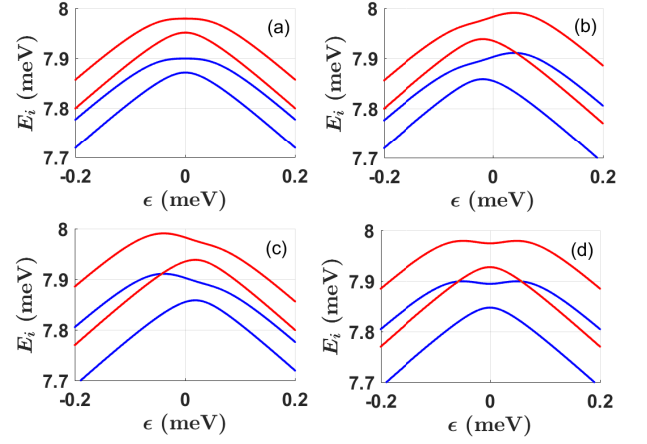}
	
	\caption{\label{fig:Eng4}(color online) Energy diagrams for different valley
		splittings: (a) $|\Delta_{L}|=|\Delta_{R}|=30\ {\rm \mu eV}$, (b)
		$|\Delta_{L}|=30\ {\rm \mu eV}$, $|\Delta_{R}|=60\ {\rm \mu eV}$,
		(c) $|\Delta_{L}|=60\ {\rm \mu eV}$, $|\Delta_{R}|=30\ {\rm \mu eV}$,
		(d) $|\Delta_{L}|=|\Delta_{R}|=60\ {\rm \mu eV}$. All the other parameters
		are the same as in Fig.~\ref{fig:FvsDLDR}.}
\end{figure}

Here we plot the energy diagrams for the four cases identified in
Fig.~\ref{fig:FvsDLDR}, with the four panels corresponding to the
four regions marked in Fig. \ref{fig:FvsDLDR}. The energy diagrams
give clear indications on how many anti-crossings occur for a certain
configuration of parameters $|\Delta_{L}|$, $|\Delta_{R}|$, and
$E_{z}$ (the valley splitting in the left and right dot, and the
Zeeman splitting).

\section{\label{sec:Supp_Suppress}Spin flip errors in a DQD}

\begin{figure}
	\includegraphics[width=1\linewidth]{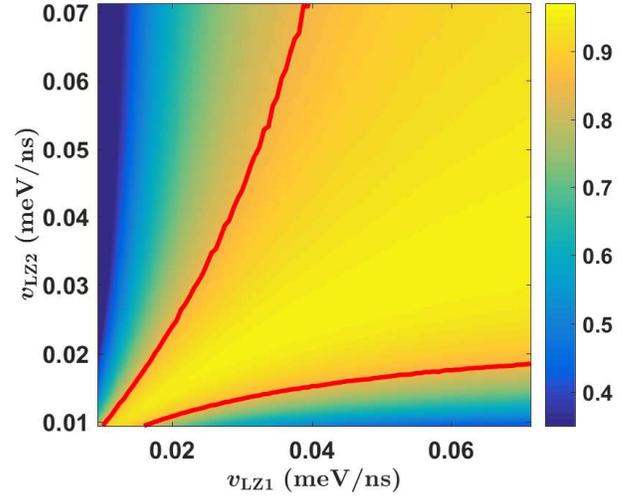}
	
	\caption{\label{fig:RecBx}(color online) Residue coherence $\frac{|\rho_{6}(1,2)|}{|\rho_{0}(1,2)|}$
		after the transport. The parameters are chosen as $E_{x}=2{\rm \mu eV}$,
		$E_{z}=40{\rm \mu eV}$, $t_{C}=15{\rm \mu eV}$. Initial state is
		$\frac{1}{\sqrt{2}}|L\rangle\otimes(|\uparrow\rangle+|\downarrow\rangle)$.}
\end{figure}

In subsection~\ref{subsec:Suppress_Spin_Flip}, we discuss a scheme to suppress spin flip errors during spin transport. The dominant spin flip error is assumed to be from spin relaxation, which is governed by a master equation
\begin{equation}
\frac{d}{dt}\rho=-\frac{i}{\hbar}[H,\rho]+\mathcal{L}_{A}(\rho)\,,\label{eq:MEQ}
\end{equation}
where $\mathcal{L}_{A}(\rho)=\frac{\gamma_{A}}{2}\left(2\sigma_{-}\rho\sigma_{+}-\sigma_{+}\sigma_{-}\rho-\rho\sigma_{+}\sigma_{-}\right)$, and $\gamma_{A}$ indicates the relaxation rate \citep{raith2011PRB,Zhao2018SR,benito2017PRB}.  It is straightforward to show that such a process will map an initial state $\rho(0)=\left[\begin{array}{cc}
\rho_{11}(0) & \rho_{12}(0)\\
\rho_{12}^{*}(0) & \rho_{22}(0)
\end{array}\right]$ to a final state
\begin{equation}
\rho(t)=\left[\begin{array}{cc}
\Gamma_{A}^{2}(t)\rho_{11}(0) & \Gamma_{A}(t)\rho_{12}(0)\\
\Gamma_{A}(t)\rho_{12}^{*}(0) & [1-\Gamma_{A}^{2}(t)]\rho_{11}(0)+\rho_{22}(0)
\end{array}\right]\,,\label{eq:ADChannel}
\end{equation}
which is what we obtained in Eq.~(\ref{eq:Relaxation}).

Spin flip errors can also come from the unwanted transitions during spin transport. Consider the example we used in subsection~\ref{subsec:Suppress_Spin_Flip}, where non-adiabatic transitions may happen at an anti-crossing.  Such a transition would cause an amplitude damping similar (but not exactly the same) to the process described by Eq.~(\ref{eq:ADChannel}). In this
case, we numerically simulate the transport process $\rho_2 \rightarrow \rho_3$ in step 2
by solving $\frac{d}{dt}\rho=-\frac{i}{\hbar}[H,\rho]$ with an initial
state $\rho_{2}$.
According to the numerical simulation, without the protection
steps 1 and 3, such a transport operation results in an 85\% residue
coherence due to the non-adiabatic transitions ``A'' and ``B''
in Fig.~\ref{fig:LZoperation}. The residue coherence $\frac{|\rho_{6}(1,2)|}{|\rho_{0}(1,2)|}$
with protections is plotted in Fig.~\ref{fig:RecBx}.
A better residue coherence is obtained in the region between the two red-solid
lines (which are contour lines for $\frac{|\rho_{6}(1,2)|}{|\rho_{0}(1,2)|}=0.85$).

The numerical simulation shows the viability of our scheme when spin flip is caused by unwanted transitions. Besides, there is a wide range to choose $v_{{\rm LZ1}}$ and $v_{{\rm LZ2}}$ to obtain better residue coherence.  Better yet, there is a wide range of sweeping speeds when the initial coherence can be completely recovered ($\frac{|\rho_{6}(1,2)|}{|\rho_{0}(1,2)|}\approx1$).

\section{\label{sec:LZ2}The recovery operation}

In this section, we analyze the recovery operation ``LZ2'' in Fig.~\ref{fig:LZoperation}~(b),  and derived the expressions given in the main text.

The recovery operation can be decomposed into three steps labeled as ``1'', ``2'', and ``3'' in Fig.~\ref{fig:LZoperation}~(b).  Let us assume that the state after the transport is a general pure state
\begin{equation}
|\psi_3\rangle=|R\rangle \otimes(c_1|\uparrow\rangle+c_2|\downarrow\rangle)\,.
\end{equation}
During step one (black arrow labeled as ``1''), the sweeping pulse is too fast to change the state, so that all anti-crossings can be approximately regarded as crossings. Therefore, after this pulse, the state is still $|\psi_4\rangle=|R\rangle \otimes({c_1|\uparrow\rangle}+{c_2|\downarrow\rangle})$.

In step two (green arrow labeled as ``2''), the state $|R,\uparrow\rangle$ will be unchanged while the state $|R,\downarrow\rangle$ is split into $|R,\downarrow\rangle$ and $|L,\uparrow\rangle$ due to the anti-crossing at A. The state $|\psi_4\rangle$ thus evolves into
\begin{equation}
|\psi_5\rangle=c_1|R,\uparrow\rangle+\sqrt{p^\prime} c_2|R,\downarrow\rangle+\sqrt{1-p^\prime} c_2|L,\uparrow\rangle\, .
\end{equation}

Step three (black arrow labeled as ``3'') is similar to step one, during which all states remain unchanged due to the fast sweeping speed. Performing a charge detection of $|R\rangle$ on the state collapses it into $|\psi_6\rangle = c_1|R,\uparrow\rangle+\sqrt{p^\prime} c_2|R,\downarrow\rangle$.  At the end, the total operation ``LZ2'' changes a state $\rho_3$ to $\rho_6$ as
\begin{equation}
\left[\begin{array}{cc}
|c_{1}|^{2} & c_{1}c_{2}^{*}\\
c_{1}^{*}c_{2} & |c_{2}|^{2}
\end{array}\right]\rightarrow \frac{1}{N^\prime} \left[\begin{array}{cc}
|c_{1}|^{2} & \sqrt{p^{\prime}}c_{1}c_{2}^{*}\\
\sqrt{p^{\prime}}c_{1}^{*}c_{2} & p^{\prime}|c_{2}|^{2}
\end{array}\right]\, ,  \label{eq:LZ2}
\end{equation}
which is exactly the transformation from Eq.~(\ref{eq:Relaxation}) to Eq.~(\ref{eq:FinalState}).  The derivation here is based on an initial pure state.  It is straightforward to show that Eq.~(\ref{eq:LZ2}) also holds for mixed states.

During the recovery LZ operation, the interdot detuning is temporarily swept all the way back to $\epsilon<0$ so that the $|R\rangle$ state is now an excited state and may suffer relaxation to the $|L\rangle$ state.  Fortunately, such a relaxation does not destroy the recovery operation, because after step ``3'' (shifting back to $\epsilon>0$), all $|L\rangle$ components will be discarded in the charge measurement.  Therefore, relaxation during the recovery operation would lower the success probability of the operation, but has no impact on the form of the final state.

In addition to the recovery protocol proposed in the main text, an alternative is to only use anti-crossing ``B'' to recover the initial state. The operation is similar to the process in Fig.~\ref{fig:LZoperation}~(a). However, the $B$-field needs to be reversed right before such an operation in order for it to lower the probability of $|\uparrow\rangle$ state.  Such a field reversal requires fast control (in nanoseconds) over magnetic field.

\bibliographystyle{apsrev4-1}
\bibliography{SiQDBIB}

\end{document}